\documentclass{vldb}

\usepackage{amsmath}
\usepackage{amssymb}

\usepackage{graphicx}
\usepackage{subfigure}
\usepackage[np,autolanguage]{numprint}
\usepackage{units}
\usepackage{balance}
\usepackage{algorithm}
\usepackage[noend]{algorithmic}

\usepackage[pdftex,pdfpagelabels=false]{hyperref} 
\hypersetup{
    unicode=false,          
    pdftoolbar=true,        
    pdfmenubar=true,        
    pdffitwindow=true,      
    pdfstartview={FitV},    
    pdftitle={On Optimizing Operator Fusion Plans for Large-Scale Machine Learning in SystemML},    
    pdfauthor={Matthias Boehm, Berthold Reinwald, Dylan Hutchison, Alexandre V. Evfimievski, Prithviraj Sen}, 
    pdfsubject={},   
    pdfcreator={},   
    pdfproducer={}, 
    pdfkeywords={}, 
    pdfnewwindow=true,      
    bookmarksnumbered=true, 
    bookmarksopen=true,     
    bookmarksopenlevel=2,   
    colorlinks=false,        
    linkcolor=red,        
    citecolor=green,         
    filecolor=black,        
    urlcolor=black          
}

\newcommand{\mat}[1]{\ensuremath{\mathbf{#1}}}
\newcommand{\card}[1]{\lvert #1\rvert}
\newcommand{\num}[1]{\numprint{#1}}

\newcommand{\s}{\unit{\,s}}

\newcommand{\bb}{\unit{\,B}}
\newcommand{\kb}{\unit{\,KB}}
\newcommand{\mb}{\unit{\,MB}}
\newcommand{\gb}{\unit{\,GB}}
\newcommand{\tb}{\unit{\,TB}}
\newcommand{\gbs}{\unit{\,GB/s}} 
\newcommand{\gflops}{\unit{\,GFLOP/s}} 

\newcommand{\COMMENTLINE}[1]{\STATE \textit{// #1}}

\newenvironment{itemize2}{\begin{itemize}\setlength{\itemsep}{2.5pt}\setlength{\parskip}{0pt}\setlength{\parsep}{0pt}}{\end{itemize}}
\newenvironment{equation2}{\begin{equation}\small}{\end{equation}\hspace{-0.1cm}}

\renewcommand{\footnotesize}{\small}
\newcommand{\eat}[1]{}
\newtheorem{definition}{Definition}
\DeclareMathOperator*{\argmin}{arg\,min}

\begin{document}

\title{On Optimizing Operator Fusion Plans\\for Large-Scale Machine Learning in SystemML}

\numberofauthors{1}
\author{
  \alignauthor 
	~\vspace{-0.7cm}\\
	Matthias Boehm\textsuperscript{1},~~~Berthold Reinwald\textsuperscript{1},~~~Dylan Hutchison\textsuperscript{2}\thanks{Work done during an internship at IBM Research -- Almaden.},\vspace{0.1cm}\\
	Alexandre V. Evfimievski\textsuperscript{1},~~~Prithviraj Sen\textsuperscript{1}\\~\\
  \affaddr{\textsuperscript{1} IBM Research -- Almaden;~~San Jose, CA, USA}\\
	\affaddr{\textsuperscript{2} University of Washington;~~Seattle, WA, USA}
}

\maketitle

\begin{abstract}
Many large-scale machine learning (ML) systems allow specifying custom ML algorithms by means of linear algebra programs, and then automatically generate efficient execution plans. In this context, optimization opportunities for fused operators---in terms of fused chains of basic operators---are ubiquitous. These opportunities include (1) fewer materialized intermediates, (2) fewer scans of input data, and (3) the exploitation of sparsity across chains of operators.
Automatic operator fusion eliminates the need for hand-written fused operators and significantly improves performance for complex or previously unseen chains of operations. However, existing fusion heuristics struggle to find good fusion plans for complex DAGs or hybrid plans of local and distributed operations. 
In this paper, we introduce an optimization framework for systematically reason about fusion plans that considers materialization points in DAGs, sparsity exploitation, different fusion template types, as well as local and distributed operations. In detail, we contribute algorithms for (1) candidate exploration of valid fusion plans, (2) cost-based candidate selection, and (3) code generation of local and distributed operations over dense, sparse, and compressed data.    
Our experiments in SystemML show end-to-end performance improvements with optimized fusion plans of up to 21x compared to hand-written fused operators, with negligible optimization and code generation overhead.
\end{abstract}

\section{Introduction}

Large-scale machine learning (ML) aims to build predictive models from large data collections \cite{CohenDDHW09} and commonly relies on data-parallel frameworks such as MapReduce \cite{DeanG04} and Spark \cite{ZahariaCDDMMFSS12} for cost-effective parallelization on commodity hardware. Large-scale ML applications range from data-intensive, traditional classification, regression, and clustering use cases, to compute-intensive matrix factorization and deep learning architectures. In this context, state-of-the-art ML systems allow data scientists to express their ML algorithms in linear algebra and statistical functions \cite{AbadiBCCDDDGIIK16,BoehmDEEMPRRSST16,HuangB013,LuoGGPJ17,RohrmannSRG17,MahoutSamsara,StonebrakerBPR11,YuSC15,YuTAMAO17}, and automatically compile efficient execution plans. This high-level specification simplifies the development of custom ML algorithms, and allows the adaptation of execution plans to different deployments as well as different data, hardware, and cluster characteristics.

\begin{figure}[!t]
  \vspace{-0.1cm}
	\centering
	\subfigure[Intermediates]{
	   \label{fig1a}\includegraphics[scale=0.4]{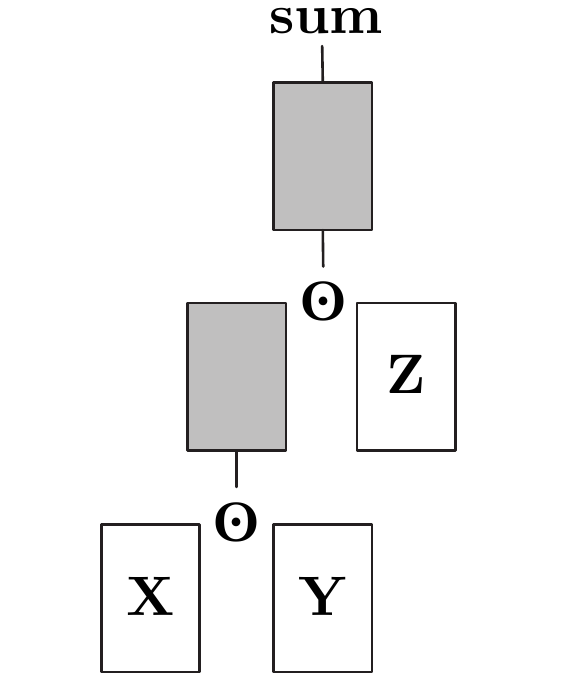}}
	\hfill	
	\subfigure[Single-Pass]{ 
	   \label{fig1b}\includegraphics[scale=0.4]{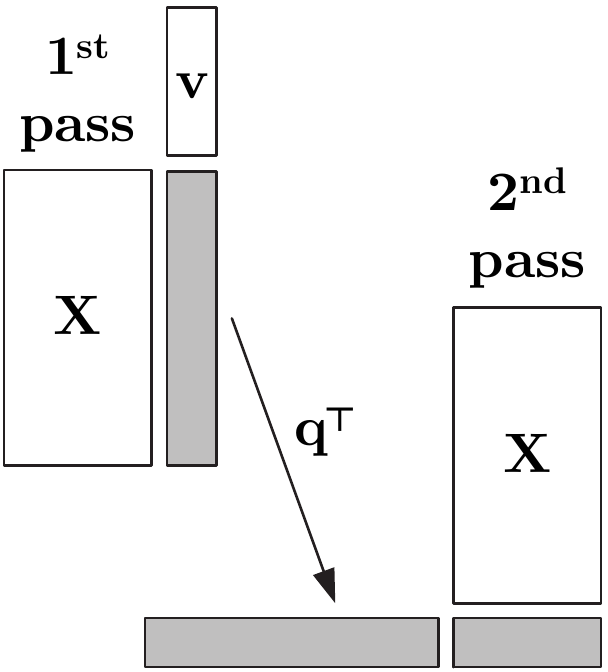}}
	\hfill	
	\subfigure[Multi-Aggregates]{
	   \label{fig1c}\includegraphics[scale=0.4]{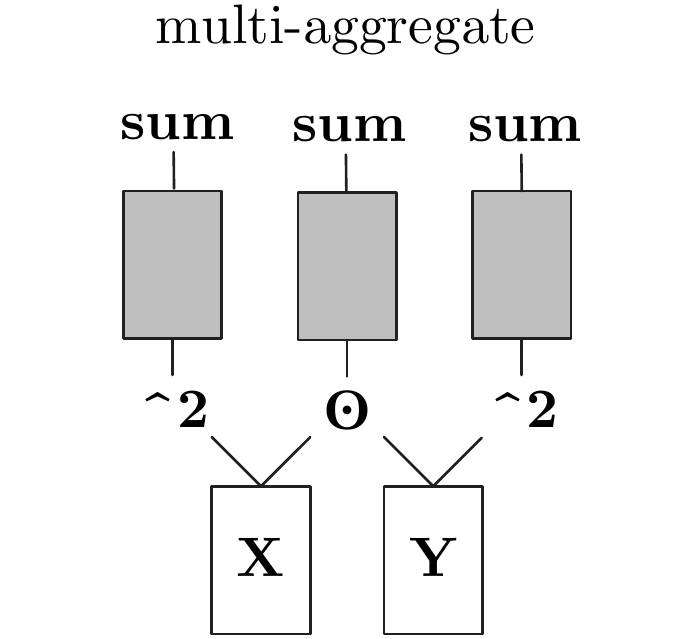}}\vspace{-0.2cm}~\\
	\subfigure[Sparsity Exploitation]{
	   \label{fig1d}\includegraphics[scale=0.4]{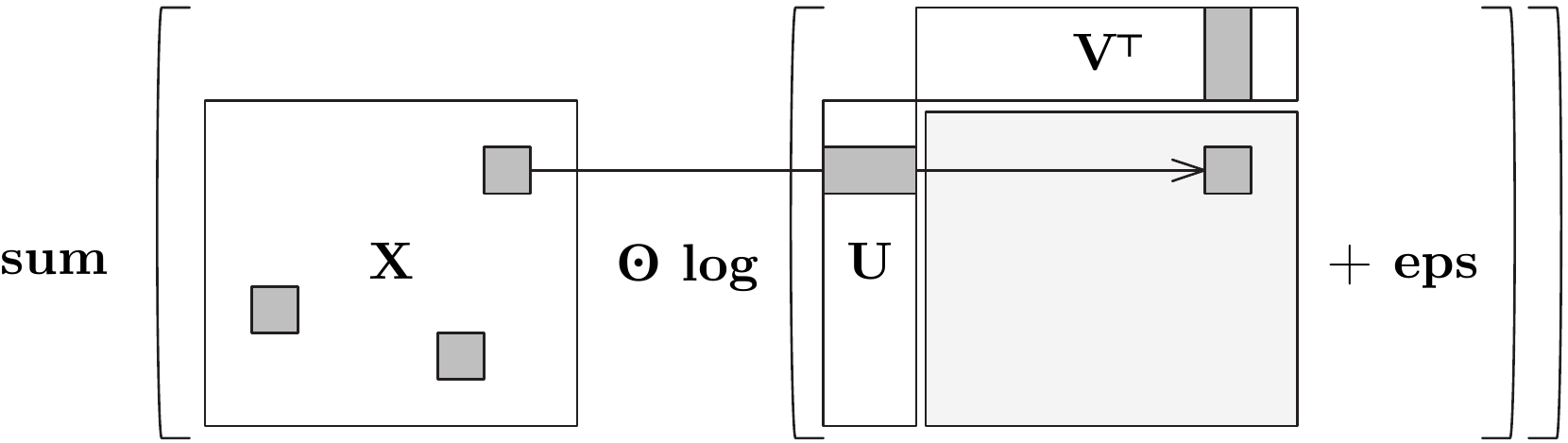}}	
   \vspace{-0.35cm}	
  \caption{\label{fig:potential}Examples of Fusion Opportunities.}
\end{figure}

\textbf{Fusion Opportunities:} The generation of execution plans has many opportunities, where fused operators---in terms of composite operators for chains of basic operators---can improve performance \cite{ElgamalLBETRS17}. Figure~\ref{fig:potential} shows major categories of fusion potential. First, fusion allows to eliminate materialized intermediates (e.g., the two intermediates in $\text{sum}(\mat{X}\odot\mat{Y}\odot\mat{Z})$ of Figure~\ref{fig1a}), whose allocation and write is often much more expensive than reading the inputs and computing the results. Second, fusion can eliminate unnecessary scans of inputs (e.g., $\mat{X}^{\top}(\mat{X}\mat{v}) \rightarrow ((\mat{X}\mat{v})^{\top}\mat{X})^{\top}$ in Figure~\ref{fig1b}) by exploiting temporal locality. Third, multiple aggregates with shared inputs (e.g., $\text{sum}(\mat{X}^2)$, $\text{sum}(\mat{X} \odot \mat{Y})$, and $\text{sum}(\mat{Y}^2)$ in Figure~\ref{fig1c}) leverage similar opportunities for DAGs (directed acyclic graphs) of multiple aggregates over common subexpressions (CSEs). Fourth, ``sparse drivers''---i.e., sparse matrices with sparse-safe operations such as multiply---allow sparsity exploitation across chains of operations (e.g., $\text{sum}(\mat{X}\odot\text{log}(\mat{U}\mat{V}^{\top}+\text{eps}))$ in Figure~\ref{fig1d}), which changes the asymptotic behavior by avoiding huge dense intermediates and unnecessary computation. 

\textbf{Existing Work on Operator Fusion:} Given the ubiquitous opportunities and high performance impact, operator fusion and code generation have received a lot of attention in the database and high performance computing literature. SystemML, for instance, uses hand-coded fused operators to eliminate intermediates \cite{HuangBTRTR15}, unnecessary scans \cite{AshariTBRCKS15}, and exploit sparsity across operations \cite{BoehmDEEMPRRSST16}. Similarly, Cumulon \cite{HuangB013} and FAQ \cite{KhamisNR16} exploit sparsity via masked operators and semijoin reductions (worst-case optimal joins), respectively. However, such approaches require custom operators that are usually limited to fixed patterns of few operators and impose large development effort for combinations of dense and sparse inputs. Automatic operator fusion addresses these issues by access-pattern-aware fusion and subsequent code generation. Example systems include BTO \cite{BelterJKS09}, OptiML \cite{SujeethLBRCWAOO11}, Tupleware \cite{CrottyGDKBCZ15,CrottyGDKCZ15}, Kasen \cite{ZhangWCMZ16}, SystemML-SPOOF \cite{ElgamalLBETRS17}, Weld \cite{abs-1709-06416,PalkarTSSAZ17}, Julia \cite{BezansonEKS17,juliagen}, TensorFlow XLA \cite{AbadiBCCDDDGIIK16,xla}, Intel Nervana Graph \cite{nervana}, and NVIDIA TensorRT \cite{tensorrt}. However, existing works mostly ignore sparse and compressed inputs, sparsity-exploiting operators, and---except for node-local optimizations in BTO~\cite{BelterJKS09} or micro-optimizations such as predication and loop tiling in Tupleware~\cite{CrottyGDKBCZ15}---do not consider the optimization of operator fusion plans. 

\textbf{A Case for Optimizing Fusion Plans:} The lack of a principled approach for optimizing fusion plans becomes increasingly problematic as code generators become more sophisticated (e.g., by covering more operations). In this context, the key challenges are DAGs of operations, overlapping access patterns, and the goal of sparsity exploitation, which create a search space that requires automatic optimization:
\begin{itemize2}
\item Materialization points (e.g., for multiple consumers),
\item Sparsity exploitation and ordering of sparse inputs,
\item Decisions on fusion patterns (e.g., template types), and
\item Constraints (e.g., memory budget and blocksizes) and costs for local and/or distributed operations.
\end{itemize2}
For example, the decision on materialization points considers redundant computation versus materialization and needs to compare an exponential number of plans. Baseline solutions are heuristics such as \emph{fuse-all} or \emph{fuse-no-redundancy}, but these heuristics struggle to find good plans for complex DAGs or hybrid plans of local and distributed operations.

\textbf{Contributions:} In this paper, we introduce a cost-based optimization framework for operator fusion plans over DAGs of linear algebra operations and describe its integration into SystemML. We formulate the optimization problem in terms of three phases---candidate exploration, candidate selection, and code generation---for which we devise novel and efficient algorithms. Our detailed technical contributions are reflected by the structure of the paper:
\begin{itemize2}
\item \emph{System Architecture:} We describe the integration into SystemML in Section~\ref{sec:sysarch}. This overview includes the compiler integration, our optimization framework, code generation plans, and their runtime integration.
\item \emph{Candidate Exploration:} In Section~\ref{sec:planexplore}, we introduce a novel bottom-up algorithm for the efficient exploration of valid partial fusion plans. We also discuss our memoization data structure and basic pruning rules. 
\item \emph{Candidate Selection:} In Section~\ref{sec:plansel}, we then present strategies for selecting the optimal candidate plans. We formulate the problem, describe heuristics, and introduce our novel cost-based plan selection, including its search space, cost model, and enumeration algorithm.
\item \emph{Experiments:} In Section~\ref{sec:eval}, we then report on extensive experiments in SystemML. These cover micro benchmarks for code generation, end-to-end performance in local and distributed environments, as well as comparisons with Julia, TensorFlow, and fusion heuristics. 
\end{itemize2}

\section{System Architecture}
\label{sec:sysarch}

In this section, we describe the architecture of our code generator (codegen) and its compiler integration into SystemML \cite{BoehmBERRSTT14,BoehmDEEMPRRSST16}. 
Our cost-based optimization framework extends the SPOOF framework \cite{ElgamalLBETRS17}, which relied on ad-hoc candidate exploration and the \emph{fuse-all} heuristic. As background, we also provide an overview of code generation plans, generated operators, and their runtime integration.

\begin{figure}[!t]
  \centering
	\includegraphics[scale=0.45]{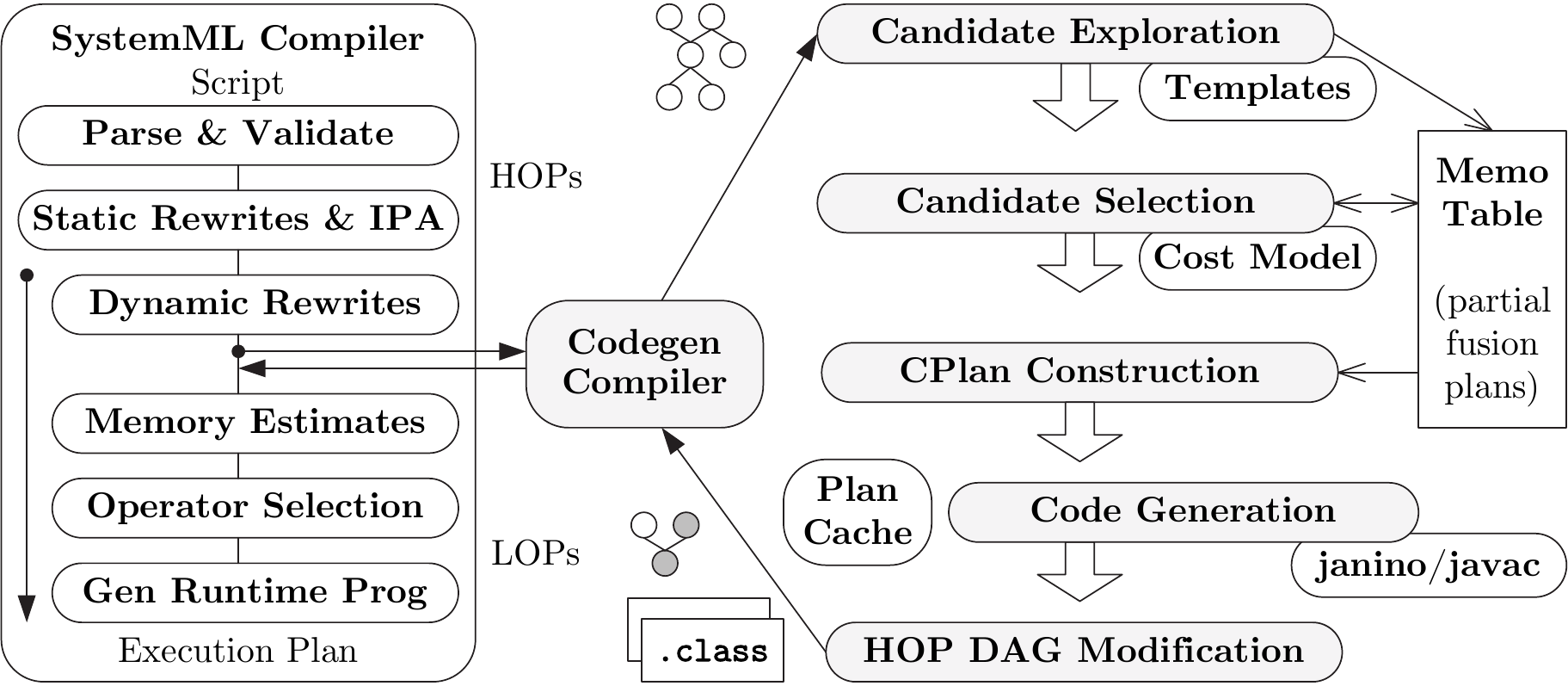}
	\vspace{-0.25cm}
	\caption{\label{fig:arch}System Architecture Overview.}
	\vspace{0.2cm}
\end{figure}

\begin{figure*}[!t]
 \centering
 \begin{minipage}{.7\textwidth}
  \subfigure[Outer Template]{
	   \label{fig3a}\includegraphics[scale=0.5]{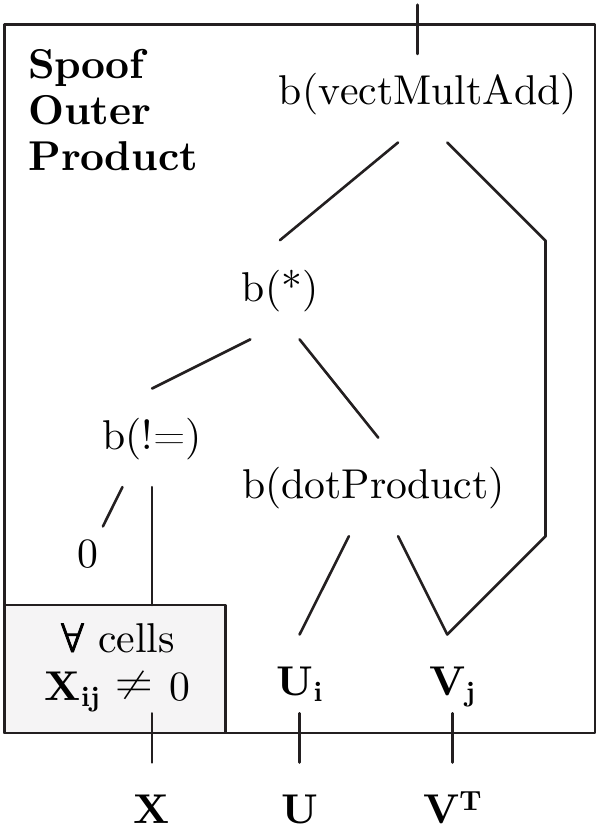}}
	\hfill	
	\subfigure[Cell Template]{
	   \label{fig3b}\includegraphics[scale=0.5]{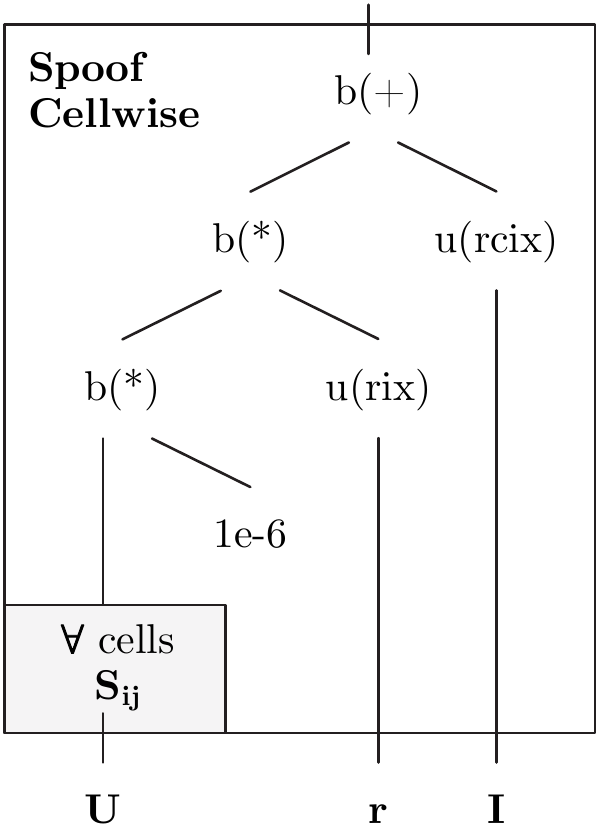}}
	\hfill	
	\subfigure[Row Template]{
	   \label{fig3c}\includegraphics[scale=0.5]{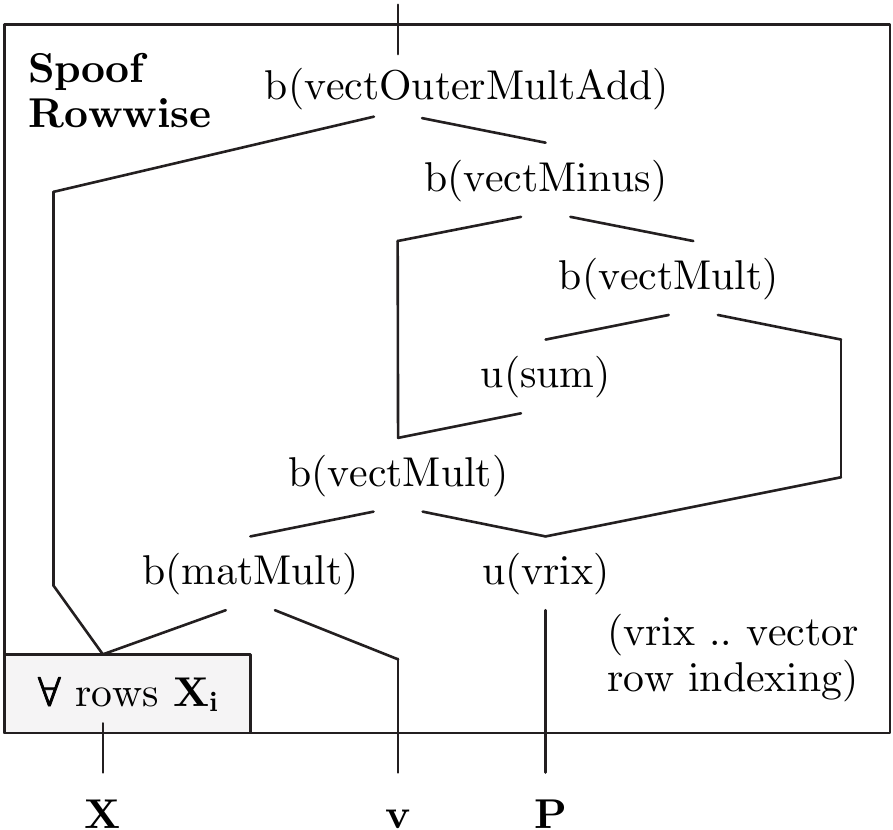}}
	\vspace{-0.45cm}
	\caption{\label{fig:cplans}Example Code Generation Plans (CPlans).}
 \end{minipage}%
 \hfill	
 \begin{minipage}{.25\textwidth}
  \vspace{0.08cm}
  \centering
  \includegraphics[scale=0.5]{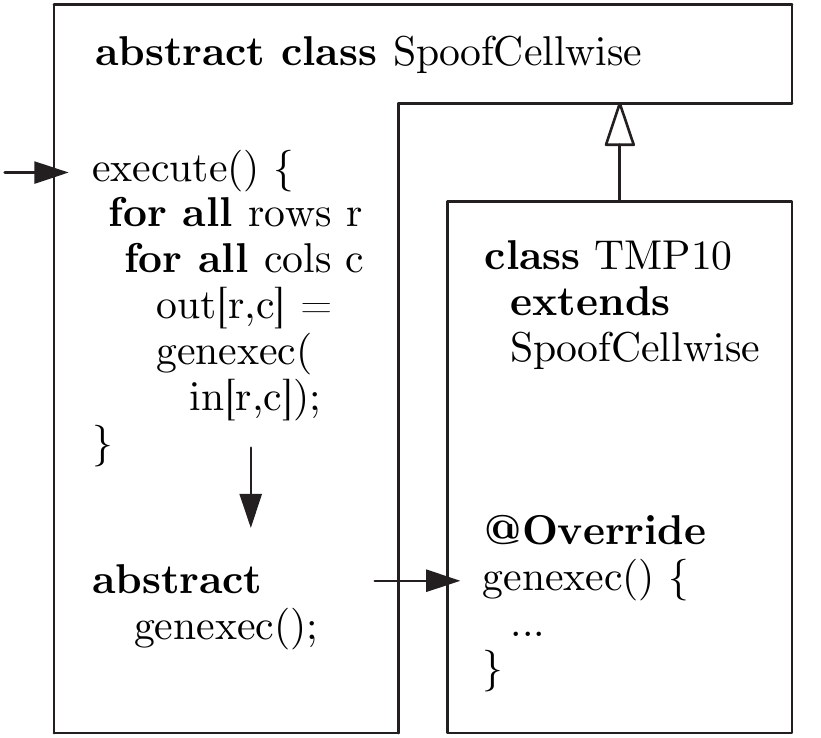}
	\caption{\label{fig:fig4}Runtime Integration of Fused Operators.}
	\vspace{0.1cm}
 \end{minipage}
 \vspace{-0.3cm}
\end{figure*}

\subsection{Compiler Integration}

SystemML provides a high-level scripting language with R-like syntax---including linear algebra, element-wise and statistical operations---to implement ML algorithms. As shown in Figure~\ref{fig:arch}, these scripts are parsed into a hierarchy of statement blocks, where blocks are delineated by control flow. Per block, we compile DAGs of high-level operators (HOPs). These DAGs are modified via static---i.e., size-independent---rewrites, and inter-procedural analysis (IPA) propagates matrix dimensions and sparsity from the inputs through the entire program. Based on this size information, we apply dynamic---i.e., size-dependent---rewrites and compute memory estimates per operation. These estimates are in turn used to select local or distributed execution types and physical operators to form an execution plan. Similar to adaptive query processing \cite{DeshpandeIR07}, SystemML recompiles HOP DAGs during runtime (from dynamic rewrites) to adapt plans for initially unknown or changing sizes \cite{BoehmBERRSTT14}.

\textbf{Codegen Compiler Integration:} Conceptually, our code generator takes---as shown in Figure~\ref{fig:arch}---the HOP DAGs after dynamic rewrites as input and produces potentially modified HOP DAGs that can include basic and fused operators. Fused operators are represented via a generic \texttt{SpoofOp}, which has a type, and references the generated and compiled class. These operators are still valid HOPs. Therefore, the remaining compilation steps of memory estimates, operator selection (e.g., local/distributed), and runtime plan generation seamlessly apply to fused operators as well. We also invoke the code generator during dynamic recompilation, which is important for many algorithms because our optimizer relies on known size information for costing and validity constraints. The actual compiler integration during initial compilation is slightly more involved. We call the code generator after runtime program generation but with access to the HOP DAGs to generate modified runtime instructions but retain the original DAG. This approach helps avoiding incomplete fusion, which loses the semantics of operations and limits fusion potential during dynamic recompilation.

\textbf{Codegen Architecture:} At a high-level, the codegen compiler comprises five well-defined compilation steps. First, on candidate exploration (Section~\ref{sec:planexplore}), we make a bottom-up pass over the HOP DAG to explore all valid partial fusion plans and store these plans in a memoization table, organized by HOPs. Second, during candidate selection (Section~\ref{sec:plansel}), we choose the optimal subset of fusion plans using a time-based cost model. Third, we construct code generation plans (CPlans, Section~\ref{sec:cplans})---which are a backend-independent basis for code generation---for all selected fusion plans. Fourth, we then recursively expand templates for these given CPlans to generate java source code for each fused operator, compile the classes and load them into the JVM. By default, we use the fast \texttt{janino} compiler \cite{janino} but also support the standard \texttt{javac} compiler. Generated fused operators are maintained in a plan cache---which identifies equivalent CPlans via hashing---to avoid redundant code generation and compilation for existing operators. Finally, we replace covered parts of the HOP DAG by the fused operators. These separate compilation steps are very valuable for debugging without compromising on fusion potential.

\subsection{Code Generation Plans}
\label{sec:cplans}

Code generation plans (CPlans) \cite{ElgamalLBETRS17} are a backend-inde\-pendent representation of fused operators and allow for recursive code generation. We generate code via a depth-first template expansion to ensure valid ordering. Such plans consist of CNodes, which are either template or basic operation nodes. Template nodes represent generic fused operator skeletons that have a specific data binding and contain a DAG of basic operations that encodes the data flow.

\textbf{Example Expressions:} We illustrate CPlans for two typical expressions with high performance impact of fusion. The first expression is part of an inner-loop update rule of ALS-CG (alternating least squares via conjugate gradient) \cite{BoehmDEEMPRRSST16}, which computes a low-rank factorization $\mat{X} \approx \mat{U} \mat{V}^{\top}$:
\begin{equation2} \label{eq:ex1}
\mat{O} = ((\mat{X}\neq0) \odot (\mat{U} \mat{V}^{\top})) \mat{V} + 10^{-6} \odot \mat{U} \odot \mat{r},
\end{equation2}
where $\odot$ denotes an element-wise multiply. Typically, $\mat{X}$ is large but sparse, and the rank (i.e., $\text{ncol}(\mat{U})$) is in the tens to hundreds. This expression requires---similar to Figure~\ref{fig1d}---a sparsity-exploiting operator to avoid computing and materializing the dense outer-product-like $\mat{U} \mat{V^{\top}}$.
The second expression is the core inner-loop operation of MLogreg (multinomial---i.e., multiclass---logistic regression):
\begin{equation2} \label{eq:ex2}
\begin{split}
\mat{Q} &= \mat{P}[~,1:k] \odot (\mat{X} \mat{v})\\
\mat{H} &= \mat{X}^{\top} (\mat{Q} - \mat{P}[~,1:k] \odot \text{rowSums}(\mat{Q})),
\end{split}
\end{equation2}
where $\mat{X}$ is the feature matrix and $k=\text{\#classes}-1$. This pattern requires---similar to Figure~\ref{fig1b}---fusion to avoid multiple passes over $\mat{X}$ and intermediates of size $\text{nrow}(\mat{X}) \times k$. 

\textbf{Code Generation:} Figure~\ref{fig:cplans} shows the three CPlans of fused operators, constructed for our example expressions. Figure~\ref{fig3a} shows the CPlan of an outer-product operator for $\mat{I} = ((\mat{X}\neq 0) \odot (\mat{U} \mat{V}^{\top})) \mat{V}$, which is sparsity-exploiting and thus improves performance proportional to the sparsity of $\mat{X}$. From this CPlan, we generate the following code: 

{\small\begin{verbatim}
 1: public final class TMP4 extends SpoofOuterProduct {
 2:  public TMP4() {super(OutProdType.RIGHT);}
 3:  protected void genexec(double a,double[] a1,int a1i, 
 4:  double[] a2,int a2i,double[] c,int ci,...,int len) {
 5:    double TMP1 = (a != 0) ? 1 : 0;
 6:    double TMP1 = dotProduct(a1, a2, a1i, a2i, len);
 7:    double TMP2 = TMP0 * TMP1;
 8:    vectMultAdd(a2, TMP2, c, a2i, ci, len);  }}
\end{verbatim}\normalsize} 

\noindent For each non-zero value $\mat{X}_{ij}$, we compute the scalar inner product $w_{ij}$ of row vectors $\mat{U}_i$ and $\mat{V}_j$, scale $\mat{V}_j$ by $w_{ij}$ and add it to the output with $\mat{I}_i \text{+=} w_{ij} \odot \mat{V}_j$, where \texttt{dotProduct} and \texttt{vectMultAdd} refer to a library of vector primitives. Figure~\ref{fig3b} shows the CPlan of an additional cell-wise operator for $\mat{I}+10^{-6}\odot \mat{U} \odot \mat{r}$, which avoids two matrix intermediates but cannot be fused into the previous Outer template due to aggregation and sparse-safeness properties. 

{\small\begin{verbatim}
 1: public final class TMP10 extends SpoofCellwise {
 2:  public TMP10() {super(CellType.NO_AGG, null, false);}
 3:  protected double genexec(double a,SideInput[] b, 
 4:  double[] scalars,..., int rix, int cix) {
 5:    double TMP5 = getValue(b[0], n, rix, cix);
 6:    double TMP6 = a * 1.0E-6;
 7:    double TMP7 = getValue(b[1], rix);
 8:    double TMP8 = TMP6 * TMP7;
 9:    double TMP9 = TMP5 + TMP8;
10:    return TMP9;  }}
\end{verbatim}\normalsize} 

\noindent Finally, Figure~\ref{fig3c} shows the row-wise CPlan of Expression~\eqref{eq:ex2}. This fused operator requires only a single pass over $\mat{X}$ by exploiting temporal locality and it avoids six large intermediates. The memory for row intermediates is managed via a preallocated ring buffer per thread (here of size 5).

{\small\begin{verbatim}
 1: public final class TMP25 extends SpoofRowwise {
 2:  public TMP25() {super(RowType.COL_AGG_B1_T,true,5);}
 3:  protected void genexecDense(double[] a, int ai, 
 4:  SideInput[] b, double[] c,..., int len) {
 5:    double[] TMP11 = getVector(b[1].vals(rix),...);
 6:    double[] TMP12 = vectMatMult(a,b[0].vals(rix),...);
 7:    double[] TMP13 = vectMult(TMP11, TMP12, 0,0,...);
 8:    double TMP14 = vectSum(TMP13, 0, TMP13.length);
 9:    double[] TMP15 = vectMult(TMP11, TMP14, 0,...);
10:    double[] TMP16 = vectMinus(TMP13, TMP15, 0,0,...);
11:    vectOuterMultAdd(a, TMP16, c, ai,0,0,...); }
12:  protected void genexecSparse(double[] avals, int[]
13:  aix, int ai, SideInput[] b, ..., int len){...}}
\end{verbatim}\normalsize} 

\textbf{Template Types:} Generalizing the previous examples, Table~\ref{tab:templates} shows all of our template types. The row-wise (Row) template binds to sparse/dense rows of a main input, a list of sparse/dense side inputs $\mathcal{Y}^{\pm}$, and a vector of scalars $\mat{s}$. Template variants represent aggregation types such as row- or column-wise aggregation, which have different implementations and allow for size propagation. Similarly, the cell-wise (Cell) and multi aggregate (MAgg) templates bind to cells $\mat{X}_{ij}$ with sparse/dense side inputs. All templates can be marked sparse-safe in which case they would only bind to non-zero rows or cells of their main input. Finally, the outer-product (Outer) template binds to non-zero cells in $\mat{X}$, rows in $\mat{U}$ and $\mat{V}$, and dense side inputs $\mathcal{Y}^{+}$. 

\begin{table}[!t]
\vspace{-0.2cm}
\centering \small \setlength\tabcolsep{6.4pt}
  \caption{\label{tab:templates}CPlan Template Types and their Variants.}
  \begin{tabular}{|c|c|c|}
	  \hline
    \textbf{Name} & \textbf{Binding} & \textbf{Variants} \\ 
    \hline
    Row& $\mat{X}_i$, $\mathcal{Y}^{\pm}$, $\mat{s}$& no\_agg, row\_agg, col\_agg, full\_agg,\\
		& & col\_t\_agg,  no/col\_agg\_B1\\
		Cell & $\mat{X}_{ij}$, $\mathcal{Y}^{\pm}$, $\mat{s}$& no\_agg, row\_agg, col\_agg, full\_agg\\
		MAgg & $\mat{X}_{ij}$,$\mathcal{Y}^{\pm}$, $\mat{s}$&full\_agg\\
		Outer & $\mat{X}_{ij} \neq 0$, $\mat{U}_i$,& left\_mm, right\_mm,\\
		& $\mat{V}^{\top}_j$, $\mathcal{Y}^{+}$, $\mat{s}$& no\_agg, full\_agg\\
		\hline
  \end{tabular}
	\normalsize
\end{table}

\textbf{Runtime Integration:} Templates refer to generic skeletons of fused operators, which are inspired by algorithmic skeletons \cite{Cole91}. Figure~\ref{fig:fig4} exemplifies this runtime integration using the Cell template. Unlike existing work \cite{BelterJKS09,CrottyGDKBCZ15,KjolstadKCLA17}, we made the conscious design decision not to generate the data access into the fused operators. Instead, the hand-coded skeleton implements the data access---depending on its sparse-safeness over cells or non-zero values---of dense, sparse, or compressed \cite{ElgoharyBHRR16} matrices and calls an abstract (virtual) \texttt{genexec} method for each value. Generated operators inherit this skeleton and only override the specific \texttt{genexec}, which yields very lean yet efficient operators. The skeleton also handles multi-threading, cache blocking, memory management, and pseudo-sparse-safe aggregations. Sharing common skeletons and vector primitives among fused operators can also reduce the instruction footprint and thus, L1 instruction cache misses, which is a known bottleneck in OLTP \cite{SirinTPA16} and scale-out workloads \cite{FerdmanAKVAJKPAF12}.  

\section{Candidate Exploration}
\label{sec:planexplore}

The exploration of candidate fusion plans aims to identify all valid partial fusion plans to provide a common input for different plan selection policies and simplify optimization. However, the exponential search space prohibits the enumeration of all possible plans. Instead, we enumerate \emph{partial fusion plans} per operator, which represent local fusion decisions. We describe (1) the representations of partial fusion plans in our central memoization table, and (2) an efficient algorithm for populating this memo table in a single pass over the HOP DAG, including pruning techniques.

\subsection{Memoization Table}

Our memoization (memo) table consists of a set of groups, where each group represents the output of an operator in the HOP DAG, i.e., a logical subexpression. Each group is identified by the operator ID, has access to its operator meta data, and contains a set of valid partial fusion plans for this operator. A partial fusion plan is called a memo table entry, and can reference other groups to represent fusion decisions. This structure is similar to \emph{groups} and \emph{group expressions} in the Cascades Optimization Framework \cite{BrunoN08,Graefe95a,ShapiroMBBFHWZWV01}, but we use it merely as a compact representation of fusion plans, which only includes operators that are amenable to fusion.

\textbf{Memo Table Entries:} A memo table entry is a tuple $(\text{type},\{i_1,..,i_k\},\text{closed})$, consisting of a template type (as introduced in Table~\ref{tab:templates}), a list of input references, and a closed type. The list of inputs corresponds to HOP inputs (i.e., data dependencies) by position, and each input is either a group reference or -1, which indicate fusion or materialized intermediates. A reference from an entry to a group implies that the group contains at least one compatible fusion plan. Finally, the close status can be open valid, open invalid (i.e., an invalid entry point), closed valid, and closed invalid.

\begin{figure}[!t]
  \centering
	\includegraphics[scale=0.5]{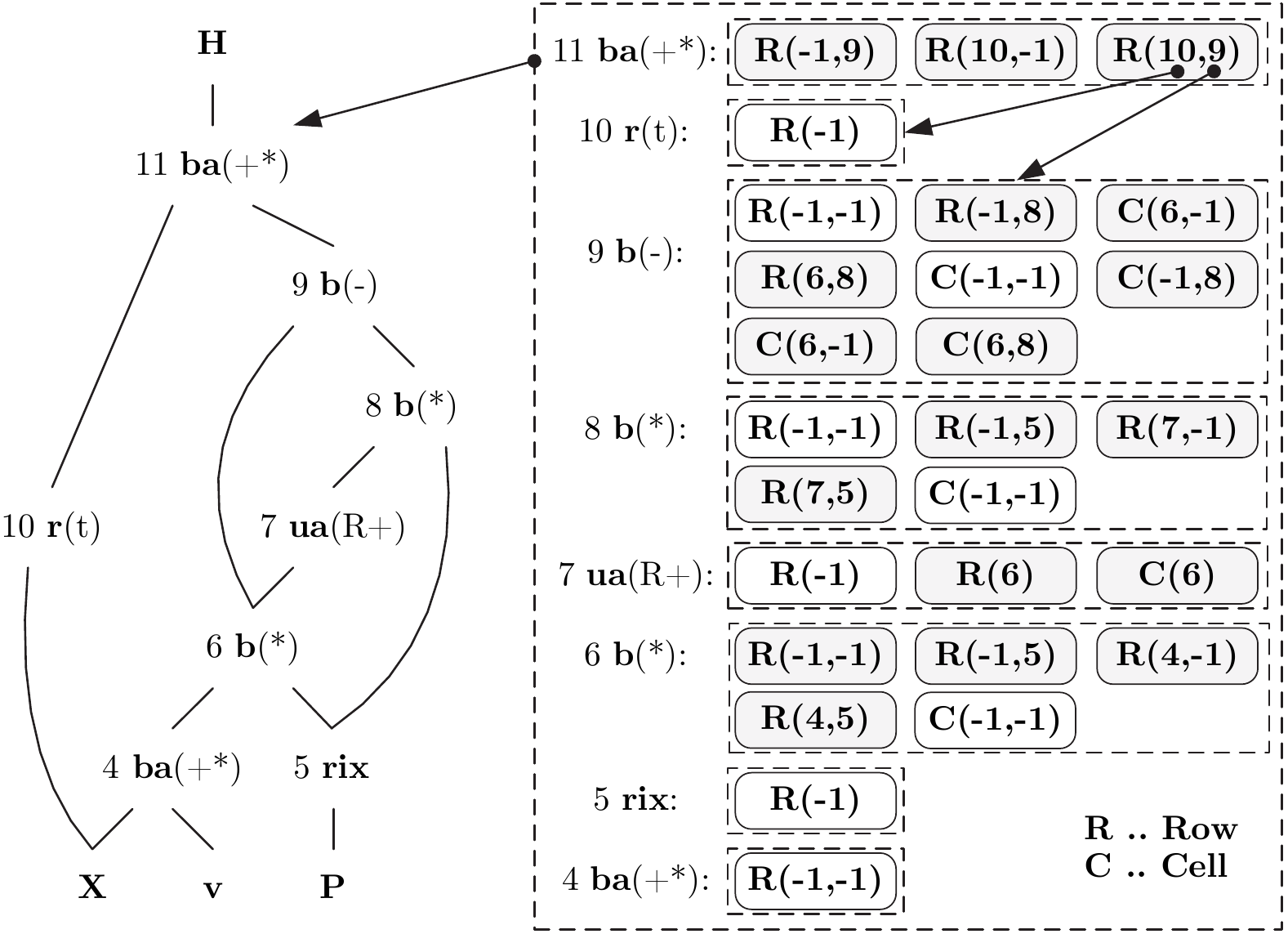}
	\vspace{-0.25cm}
	\caption{\label{fig:memo}Example Memo Table (w/ basic pruning).}
\end{figure}

\textbf{Example:} We use Expression~\eqref{eq:ex2} from Section~\ref{sec:cplans} to illustrate the structure of our memo table. Figure~\ref{fig:memo} shows the HOP DAG and the related memo table after candidate exploration and pruning (described in Section~\ref{sec:ofmc}). All eight operators are represented by groups in the memo table. The group 11 refers to the final matrix multiplication (binary aggregate \texttt{ba(+*)}), and consists of three memo table entries of type Row. These entries encode fusion alternatives: (1) fuse right R(-1,9), (2) fuse left R(10,-1), and (3) fuse both R(10,9). Instead of encoding all alternative subplans along inputs, we only reference the input groups. This memo table then allows for simple costing and fusion by traversing the HOP DAG top down, probing for fusion plans, traversing group references, and determining the input HOPs, from where this process repeats until we reach the leaf HOPs. 

\subsection{Open-Fuse-Merge-Close Exploration}
\label{sec:ofmc}

Given a HOP DAG and an empty memo table, we aim to efficiently discover all valid partial fusion plans. We introduce a bottom-up algorithm that is template-oblivious and populates the memo table in a single pass over the DAG.

\textbf{OFMC Template Abstraction:} As the basis of our candidate exploration algorithm, we define the \emph{open-fuse-merge-close} (OFMC) template abstraction: 
\begin{itemize2}
\item \texttt{open(Hop h):} Indicates if a new fused operator of this template can be started at HOP $h$, covering its operation and reading materialized inputs. For example, the condition of an Outer template is an outer-product-like matrix multiplication with size constraints.
\item \texttt{fuse(Hop h, Hop in):} Indicates if an open fused operator (of this template) at the input HOP $in$ can be expanded to its consumer HOP $h$. For example, a Cell template can fuse valid unary, binary, or ternary operations, valid aggregations, and inner products.
\item \texttt{merge(Hop h, Hop in):} Indicates if an open fused operator (of this template) at the consumer HOP $h$ can be expanded to its input HOP $in$, i.e., if it can merge with fused operators at the input. An example is the merge of Cell templates into Row templates.
\item \texttt{close(Hop h):} Indicates the close status of the template after the HOP $h$ and its validity. For example, any aggregation closes a Cell template (as valid or invalid), whereas only column-wise or full aggregations close a Row template. Outer templates are also validated for the existence of sparsity exploiting operators.
\end{itemize2}
The major benefit of this OFMC abstraction is the separation of template-specific conditions from the HOP DAG traversal and the population of the memo table.

\begin{algorithm}[!t] 
\small
  \caption{OFMC Explore (recursive)}\label{alg:explore}
		\begin{algorithmic}[1] 
		\REQUIRE{An operator $g_i$ of DAG $\mathcal{G}$, memo table $\mathcal{W}$}
		\ENSURE{A populated memo table $\mathcal{W}$}
		\COMMENTLINE{Memoization of processed operators -- -- -- -- -- -- -- -- -- -- --} \label{a1:1}
		\IF{$\exists \mathcal{W}[g_i] \vee g_i \in \mathcal{W}[\star]$} 
			\RETURN $\mathcal{W}$ \label{a1:3}
    \ENDIF
		\COMMENTLINE{Recursive candidate exploration -- -- -- -- -- -- -- -- -- -- -- -- --} \label{a1:4}
		\FORALL[for all operator inputs]{$j$ \textbf{in} 1 \textbf{to} $\card{g_i}$}
			\STATE \textsc{ofmcExplore}($g_{j}$, $\mathcal{W}$)  \label{a1:6}
		\ENDFOR
		\COMMENTLINE{Open initial operator plans  -- -- -- -- -- -- -- -- -- -- -- -- -- -- --} \label{a1:7}
		\FORALL[for all template types]{$t \in T$}
			\IF[probe opening condition]{$t.\textsc{open}(g_i)$} 
				\STATE $\mathcal{W}[g_i] \leftarrow \textsc{createPlans}(g_i, \textbf{null}, t)$ \label{a1:10}
			\ENDIF
		\ENDFOR
		\COMMENTLINE{Fuse and merge operators plans -- -- -- -- -- -- -- -- -- -- -- -- --} \label{a1:11}
		\FORALL[for all operator inputs]{$j$ \textbf{in} 1 \textbf{to} $\card{o_i}$}
			\FORALL[for all distinct templates]{$t$ \textbf{in} $\mathcal{W}[g_j]$}
				\IF{$t.\textsc{fuse}(g_i, g_j)$}
				  \STATE $\mathcal{W}[g_i] \leftarrow \mathcal{W}[g_i] \cup \textsc{createPlans}(g_i, g_j, t)$ \label{a1:16}
				\ENDIF
			\ENDFOR
		\ENDFOR
		\COMMENTLINE{Close operator plans if required -- -- -- -- -- -- -- -- -- -- -- -- --} \label{a1:17}
		\FORALL[for all memo entries]{$me$ \textbf{in} $\mathcal{W}[g_i]$}
		  \STATE $me.closed \leftarrow t(me.type).\textsc{close}(g_i)$
			\IF[closed invalid]{$me.closed < 0$}
				\STATE $\mathcal{W}[g_i] \leftarrow \mathcal{W}[g_i] \setminus me$ \label{a1:22}
			\ENDIF
		\ENDFOR	
    \COMMENTLINE{Prune redundant plans and memoize -- -- -- -- -- -- -- -- -- --} 
		\STATE \textsc{pruneRedundant}($\mathcal{W}$, $g_i$) \label{a1:24}
		\STATE $\mathcal{W}[\star] \leftarrow \mathcal{W}[\star] \cup g_i$
		\RETURN $\mathcal{W}$
  \end{algorithmic}
\normalsize	
\end{algorithm}

\textbf{The OFMC Algorithm:} Based on the memo table and OFMC abstraction, we introduce the OFMC exploration algorithm (Algorithm~\ref{alg:explore}). This algorithm is called recursively, in a depth-first manner to populate the memo table bottom-up. First, we check for already processed operators---indicated by existing groups or marked operators---(lines~\ref{a1:1}-\ref{a1:3}) to avoid redundant exploration if nodes are reachable over multiple paths. Second, we recursively explore all input operators (lines~\ref{a1:4}-\ref{a1:6}) because these input data dependencies constitute potential fusion references. Third, we explore all templates for valid opening conditions at the current operator (lines~\ref{a1:7}-\ref{a1:10}). In case of a valid opening condition, we add this memo entry and enumerate merge plans with \textsc{createPlans}. This merging is important to cover scenarios such as $\mat{X}^{\top} (\mat{y} \odot \mat{z})$, where the matrix-vector multiplication with $\mat{X}$ opens a Row template, which can also merge Cell templates over $\mat{y} \odot \mat{z}$. Third, we fuse and merge existing partial fusion plans from the operator inputs to the current operator (lines~\ref{a1:11}-\ref{a1:16}). This step entails iterating over all distinct template types of all inputs and probing pair-wise fusion conditions. In case of a valid condition, we again call \textsc{createPlans}, 
which constructs a memo table entry for the fused operator, and enumerates all \emph{local} plan combinations for inputs that satisfy the pair-wise merge condition. This entire plan set is then added to the group of the current operator. Fourth, we check all group entries for closing conditions (lines~\ref{a1:17}-\ref{a1:22}). Entries which satisfy the closing condition of their templates are either removed (invalid) or marked as closed (valid), while all other entries remain open.

\textbf{Pruning Techniques:} Finally, we prune duplicates and valid closed entries without group references (line~\ref{a1:24}). For example, the group \texttt{7 ua(R+)} in Figure~\ref{fig:memo} does not contain C(-1) because a \texttt{rowSums} closes the Cell template, which would cover only a single operator. In addition, there are advanced techniques that exploit characteristics of candidate selection policies. For instance, a policy that only considers materialization points with multiple consumers allows pruning dominated plans. A memo entry is dominated if all its references point to operators that are consumed once, and there is another entry (of the same type) whose reference list is a strict superset. For example, in Figure~\ref{fig:memo}, R(10,9) dominates R(10,-1) but R(6,8) does not dominate R(-1,8) because group 6 has multiple consumers. However, we prune dominated plans only for selection heuristics.

\textbf{Algorithm Analysis:} Overall our algorithm has linear time and space complexity in the number of operators. Memoization ensures that we visit each operator exactly once and the OFMC conditions apply only locally to an operator and its inputs. These conditions still have access to the hops and thus the entire DAG but this flexibility is only exploited in rare exceptions such as recognizing \texttt{t(cumsum(t(X)))} as a row operation. For each operator $g_i$, we enumerate up to $O(2^{\card{g_i}} \cdot \card{T})$ memo entries, but the supported $\card{T}=4$ templates and ternary basic operators (i.e., $\card{g_i}=3$), give us an upper bound of $32n$ plans, and works very well in practice. 

\section{Candidate Selection}
\label{sec:plansel}

Given a memo table of partial fusion plans, candidate selection aims to choose the optimal subset of non-conflicting partial fusion plans. We describe the problem and cost model, as well as introduce our cost-based enumeration algorithm \textsc{MPSkipEnum}. The basic ideas are to (1) partition the set of partial fusion plans into independent groups, (2) restrict the search per group to interesting materialization points, (3) linearize the resulting exponential search space, and (4) enumerate and cost plans with skipping of search space areas that can be pruned based on cost or structure.

\subsection{Problem Formulation and Heuristics}

Overall, we aim to find the cost-optimal set of fusion plans with the optimization scope of a single HOP DAG at-a-time and hybrid runtime plans that might include single-node and distributed operations. We define this problem as follows: 

\begin{definition}\textbf{Candidate Selection Problem:} Given an operator DAG $\mathcal{G}$, and a set of partial fusions plans $P$, find the set of optimal, non-conflicting fusion plans $P^{\star}$ that applied to $\mathcal{G}$ minimizes costs $C$ with
\begin{equation2}
P^{\star} = \argmin_{p \subseteq P} C(\mathcal{G}, p)\mbox{~~}s.t.\mbox{~~} Z \vDash p,
\end{equation2} 
where $Z$ is a set of constraints such as memory budgets and block size restrictions that any plan $p \in P^{\star}$ must satisfy. Fusion plans are conflicting if they are connected via fusion references but their template types are incompatible.
\end{definition}

\textbf{Conditional Constraints:} Some constraints $z \in Z$ are conditional on the resulting execution type (local vs distributed), which is decided based on memory requirements but these memory estimates in turn depend on the number of fused inputs. For example, a Row template has no constraints for single-node operations, but it has a blocksize constraint $z: \text{ncol(\mat{X})}\leq B_c$ for distributed operations because the fused operator requires access to entire rows.

\textbf{Selection Heuristics:} Common baseline solutions to this fusion problem for DAGs are the following heuristics:
\begin{itemize2}
\item \emph{Fuse-all} aims at maximal fusion, which leads to redundant compute on CSEs. This heuristic is similar to lazy evaluation in Spark \cite{ZahariaCDDMMFSS12}, delayed arrays in Repa~\cite{KellerCLJL10}, and code generation in SPOOF \cite{ElgamalLBETRS17}. 
\item \emph{Fuse-no-redundancy} takes another extreme of fusion without redundant compute, which leads to materializing all intermediates with multiple consumers. This heuristic is similar to caching policies in Emma~\cite{AlexandrovKKSTK15}.
\end{itemize2}
In the following, we use these heuristics as baselines but focus solely on finding the cost-optimal set of fusion plans. 

\subsection{Plan Partitions and Interesting Points}

As a preparation step for plan enumeration, we analyze the set of partial fusion plans $P$, given in the populated memo table. This analysis---which is based on the maximal DAG of fusion references---includes determining independent partitions, as well as root nodes, input nodes, and interesting materialization points per partition. 

\begin{figure}[!t]
  \centering
	\includegraphics[scale=0.49]{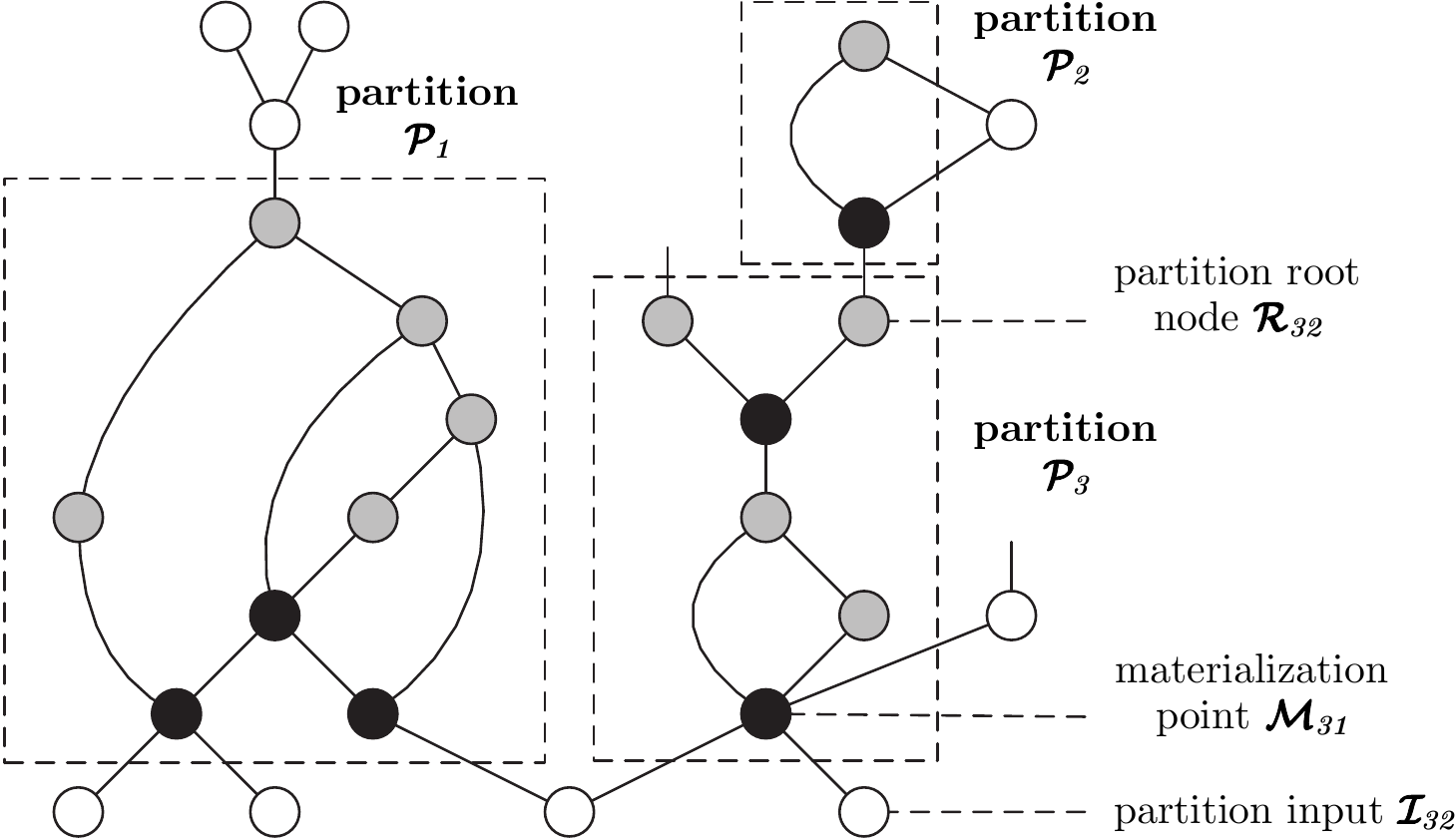}
	\vspace{-0.25cm}
	\caption{\label{fig:partition}Example Plan Partitions and Terminology.}
	\vspace{-0.1cm}
\end{figure}

\textbf{Plan Partitions:} We define the \emph{plan partitions} $\mathcal{P}$ of $P$ as its connected components in terms of fusion references. Hence, nodes of separate partitions are not reachable via fusion. This property allows optimizing and costing the partitions independently, which can significantly reduce the search space. Figure~\ref{fig:partition} shows an extended example of such a partitioning. Gray and black nodes represent HOPs with  fusion plans. In this example, we have three independent partitions, of which Partition 2 and 3 are adjacent. Such an adjacent partitioning can originate, e.g., from a column aggregation like \texttt{colSums}, which closes all templates. In addition, we define the following terminology. First, \emph{root nodes} $\mathcal{R}_i$ of a partition $\mathcal{P}_i$ (with $\mathcal{R}_i \subseteq \mathcal{P}_i$) are nodes that are never referenced from $v \in \mathcal{P}_i$. These root nodes are the entry points for partition analysis and costing. Second, \emph{input nodes} $\mathcal{I}_i$ of a partition $\mathcal{P}_i$ (with $\mathcal{I}_i \cap \mathcal{P}_i = \emptyset$) are nodes whose output is read by any node $g \in \mathcal{P}_i$. Third, \emph{materialization points} $\mathcal{M}_i$ (black nodes in Figure~\ref{fig:partition}) of a partition $\mathcal{P}_i$ (with $\mathcal{M}_i \subseteq \mathcal{P}_i \wedge \mathcal{M}_i \cap \mathcal{R}_i = \emptyset$) are nodes with multiple consumers. These materialization points are interesting for plan choices because multiple consumers can lead to redundant compute. 

\textbf{Interesting Points:} Generalizing the notion of materialization points, we define the search space per partition in terms of its \emph{interesting points} $\mathcal{M}^{\prime}_i$. These interesting points represent boolean fusion decisions and our optimizer considers the exponential space of $2^{\card{\mathcal{M}^{\prime}_i}}$ plan assignments. In detail, we collect two types of interesting points:
\begin{itemize2}
\item \emph{Materialization point consumers} $(g \rightarrow \mathcal{M}_{ij})$ with $g\in\mathcal{P}_i$ are considered individually. This fine-grained reasoning per data dependency is important for overlapping fused operators to avoid forcing the read of a materialized intermediate despite this intermediate being already available inside the fused operator.
\item \emph{Template switches} are interesting even without referencing an $\mathcal{M}_{ij}$. A switch is defined as a dependency $(g_i \rightarrow g_j)$ where $\mathcal{W}[g_j]$ contains template types that are not in $\mathcal{W}[g_i]$. Considering these switches is important for patterns like $\mat{Y} + \mat{X} \odot \mat{U}\mat{V}^{\top}$ where a Cell template fuses $\mat{Y} + \mat{X} \odot \text{TMP}$ and hence destroys the sparsity-exploiting Outer template $\mat{X} \odot \mat{U}\mat{V}^{\top}$. 
\end{itemize2}
If an interesting point $(g_i \rightarrow g_j)$---i.e., a data dependency---is assigned true, then all partial fusion plans with a reference from $g_i$ to $g_j$ are considered invalid and ignored for costing. After optimization, we simply remove all invalid plans.

\subsection{Cost Model}

Given a plan assignment $\mat{q}$ of interesting points (i.e., a boolean vector of materialization decisions), we compute the costs $C(\mathcal{P}_i | \mat{q})$ of the entire plan partition $\mathcal{P}$ with an analytical cost model for DAG-structured fusion plans including sparsity-exploitation and redundant computation as follows:
\begin{equation2}
C(\mathcal{P}_i | \mat{q}) = \sum_{p \in \mathcal{P}_i | \mat{q}} \left(\hat{T}^w_p + \max(\hat{T}^r_p, \hat{T}^c_p)\right).
\vspace{-0.15cm}
\end{equation2}
where $p$ is a basic or fused operator defined by $\mat{q}$ and $\hat{T}^w_p$, $\hat{T}^r_p$, and $\hat{T}^c_p$ are estimates for its write, read, and computation times. The read and write time estimates are derived from the size of inputs and outputs, normalized by peak read and write memory bandwidth. For example, reading a $100\text{M} \times 10$ dense input matrix at $32\gbs$ peak read bandwidth, gives us a time estimate of $\hat{T}^r_p=1\text{G} \cdot 8\bb / 32\gbs = 0.25\s$. Similarly, the compute time is derived from the number of required floating point operations and peak compute bandwidth. We take $\max(\hat{T}^r_p, \hat{T}^c_p)$ to account for overlapping read and compute costs, while adapting to I/O- and compute-bound operations. Sparsity-exploiting operators simply scale these estimates down by the sparsity of the main input.

\textbf{Cost Computation via Cost Vectors:} The costs of a partition $C(\mathcal{P}_i | \mat{q})$ are computed recursively with $\textsc{getPlanCost}(\mat{q}, \mathcal{P}_i, \mat{c}_p)$ starting from its roots $\mathcal{R}_i$. Shared reads and CSEs are captured via \emph{cost vectors} $\mat{c}_p$ per fused operator. We call \textsc{getPlanCost} at $\mathcal{R}_i$ without cost vectors, indicating that a new fused operator can be opened. At each operator, we either open or extend a fused operator and query the memo table accordingly for the best fusion plan regarding template type and fusion references. If the fusion plan has a reference to an input, we cost this input with the cost vector; otherwise, we add the operator as an input and get the subplan's cost without the cost vector. Once we recursively processed all inputs of an opened operator, we compute and add its costs using $\mat{c}_p$, i.e., its output size, compute workload, and sizes of disjoint inputs. Non-partition consumers of intermediates of a fused operator are handled by calling \textsc{getPlanCost} again without $\mat{c}_p$. Memoizing pairs of operators and cost vectors enables us to return zero costs if an operator is reachable over multiple paths with materialized output or within a fused operator, while accounting for redundant compute of overlapping operators.

\subsection{Enumeration Algorithm \textsc{MPSkipEnum}}

\begin{figure}[!t]
	\centering \vspace{-0.2cm}
	\hfill
	\subfigure[Search Space with Skip-Ahead]{
	   \label{fig7}\includegraphics[scale=0.5]{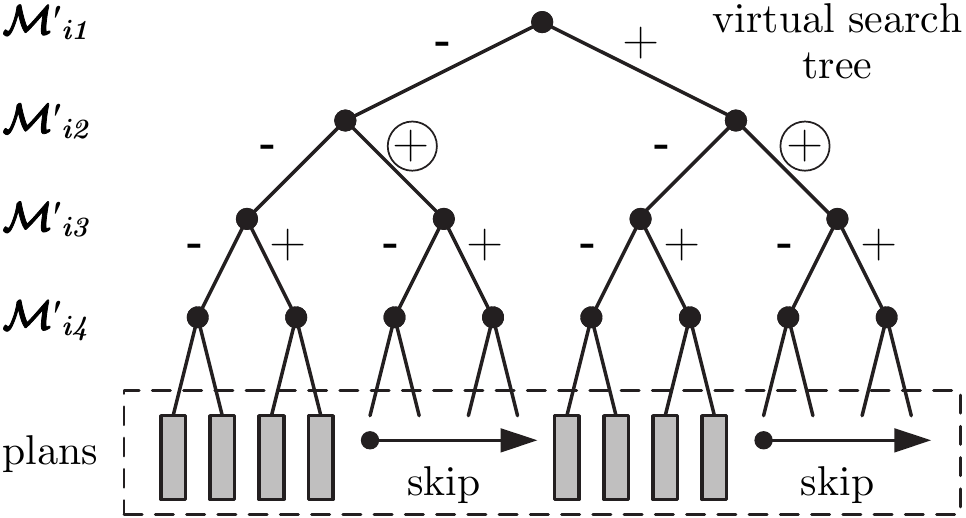}}
	\hfill	
	\subfigure[Cut Set]{
	   \label{fig8}\includegraphics[scale=0.5]{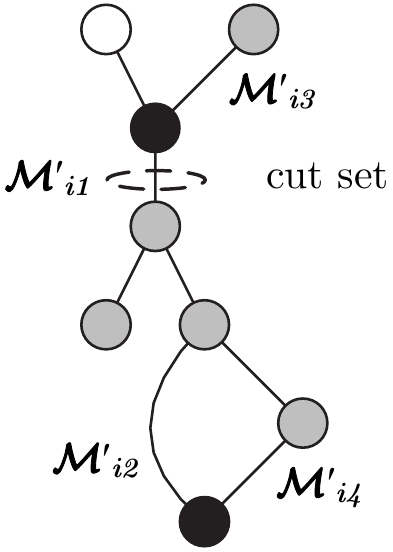}}
	\hfill		
	\vspace{-0.3cm}
  \caption{\label{fig:searchspace}Example Search Space and Cut Set.}
\end{figure}

Given an independent fusion partition $\mathcal{P}_i$, its interesting points $\mathcal{M}^{\prime}_i$, and the described cost model, we aim to find the optimal assignment $\mat{q}^{\star}$ that minimizes costs. We introduce the remarkably simple \textsc{MPSkipEnum} algorithm that linearizes the exponential search space, enumerates and costs plans with the ability to skip entire areas of the search space using cost-based and structural pruning techniques. 

\textbf{Basic Enumeration:} Algorithm~\ref{alg:skipenum} shows the basic enumeration approach. We iterate over the linearized search space of all $2^{\card{\mathcal{M}^{\prime}_i}}$ plans (lines~\ref{a3:2}-\ref{a3:20}), create a boolean plan assignment $\mat{q}$ that represents positive materialization decisions (line~\ref{a3:3}), cost the plan with \textsc{getPlanCost} (line~\ref{a3:17}), and maintain the best plan $\mat{q}^{\star}$ and its costs $\overline{C}$ (lines~\ref{a3:18}-\ref{a3:20}) as we scan through the search space. Figure~\ref{fig7} shows an example search space of $\card{\mathcal{M}^{\prime}_i}=4$ interesting points and its 16 plans. This basic enumeration approach is simple and has no space requirements. However, evaluating the exponential number of plans quickly becomes infeasible as $\card{\mathcal{M}^{\prime}_i}$ increases. Therefore, we apply two high-impact pruning techniques.

\textbf{Cost-Based Pruning (lines~\ref{a3:11}-\ref{a3:15}):} The linear scan over the search space maintains the costs of the best plan $\overline{C}$, which is an upper bound that is monotonically decreasing. By computing a lower bound $\underline{C}$ of unseen plans, we can safely prune these plans whenever $\underline{C} \geq \overline{C}$, which is very effective in practice (see Section~\ref{sec:overhead}). We compute $\underline{C}$ from static and plan-dependent costs. First, the static costs $\underline{C}_{\mathcal{P}_i}$ are the costs of reading partition inputs $\mathcal{I}_i$, minimal computation (no redundancy, and sparsity exploitation), and writing partition roots $\mathcal{R}_i$. Second, we add the minimum materialization costs of the plan $\mat{q}$ (with \textsc{getMPCost}), where each distinct target in $\mathcal{M}^{\prime}_{i}$ dependencies requires at least one write and read. Furthermore, our search space is linearized from negative to positive assignments. This layout is crucial for the effectiveness of pruning. Evaluating the plan with maximal fusion---i.e., the \emph{fuse-all} heuristic---first, yields a good upper bound $\overline{C}$ from the beginning. Additionally, this layout allows for skipping larger subareas of the search space. Figure~\ref{fig7} shows an example, where $\mathcal{M}^{\prime}_{i2}$ is set to true ($\oplus$). If the plan, with $\mathcal{M}^{\prime}_{i3}$ and $\mathcal{M}^{\prime}_{i4}$ set to false, already exceeds $\overline{C}$, we can safely prune the entire subtree because any other plan only adds materialization costs to $\underline{C}$. In this case, we compute the number of skipped plans (line~\ref{a3:14}) as $2^{\card{\mathcal{M}^{\prime}_i}-x-1}$, where $x=\text{lastIndexOf}(\mat{q},\textbf{true})$. Furthermore, we also leverage $\overline{C}$ for partial costing in $\textsc{getPlanCost}$, where we stop costing as soon as the partial plan costs exceed $\overline{C}$.

\begin{algorithm}[!t] 
\small
  \caption{Materialization Point Skip Enumerate}\label{alg:skipenum}
		\begin{algorithmic}[1] 
		\REQUIRE{memo table $\mathcal{W}$, plan partition $\mathcal{P}_i$, reachability graph $RG$,\\~~~~~~interesting points $\mathcal{M}^{\prime}_i$, offset $\text{off}$}
		\ENSURE{The best plan $\mat{q}^{\star}$}
		\STATE $\mat{q}^{\star} \leftarrow \textbf{null}$;~~~$\overline{C} \leftarrow \infty$
		\FORALL{$j$ \textbf{in} 1 \textbf{to} $2^{\card{\mathcal{M}^{\prime}_i}-\text{off}}$} \label{a3:2}
		   \STATE $\mat{q} \leftarrow \textsc{createAssignment}(\card{\mathcal{M}^{\prime}_i}-\text{off}, \text{off}, j)$ \label{a3:3}
			 \STATE $pskip \leftarrow 0$
			 \COMMENTLINE{pruning via skip-ahead  -- -- -- -- -- -- -- -- -- -- -- -- -- -- -- --}
			 \IF[structural]{$RG \neq \textbf{null} \wedge \textsc{isCutSet}(RG, \mat{q})$} \label{a3:6}
			    \STATE $pskip \leftarrow \textsc{getNumSkipPlans}(RG, \mat{q})$ 
					\STATE $S \leftarrow \textsc{getSubProblems}(RG, \mat{q})$
					\FORALL{$k$ \textbf{in} 1 \textbf{to} $\card{S}$} \label{a3:9}
					   \STATE $\mat{q}[S_k.\mat{ix}] \leftarrow \textsc{MPSkipEnum}(\mathcal{W},\mathcal{P}_i,\textbf{null},S_k.\mat{m},S_k.\text{off})$\label{a3:10}
		      \ENDFOR
			 \ELSE [cost-based] \label{a3:11}
					\STATE $\underline{C} \leftarrow \underline{C}_{\mathcal{P}_i} + \textsc{getMPCost}(\mathcal{W}, \mathcal{P}_i, \mathcal{M}^{\prime}_i, \mat{q})$
					\IF{$\underline{C} \geq \overline{C}$}
			       \STATE $j \leftarrow j + \textsc{getNumSkipPlans}(\mat{q}) - 1$ \label{a3:14}
				     \STATE \textbf{continue} \label{a3:15}
			    \ENDIF
			 \ENDIF
			 \COMMENTLINE{plan costing and comparison  -- -- -- -- -- -- -- -- -- -- -- -- --}
			 \STATE $C \leftarrow \textsc{getPlanCost}(\mathcal{W}, \mathcal{P}_i, \mathcal{M}^{\prime}_i, \mat{q}, \overline{C})$ \label{a3:17}
			 \IF{$\mat{q}^{\star} = \textbf{null} \vee C < \overline{C}$} \label{a3:18}
			    \STATE $\mat{q}^{\star} \leftarrow \mat{q}$;~~~$\overline{C} \leftarrow C$ \label{a3:19}
			 \ENDIF
			 \STATE $j \leftarrow j + pskip$ \label{a3:20}
		\ENDFOR
		\RETURN $\mat{q}^{\star}$ 
		\end{algorithmic}
\normalsize	
\end{algorithm}
 
\textbf{Structural Pruning (lines~\ref{a3:6}-\ref{a3:10}):} Similar to state-of-the-art join enumeration algorithms \cite{MoerkotteN06, MoerkotteN08}, we exploit the graph structure of a partition $\mathcal{P}_i$ and its interesting points $\mathcal{M}^{\prime}_i$ for additional pruning. The key observation is that materialization points can---predicated on their assignment and the graph structure---create independent sub-problems because these points act as fusion barriers. Figure~\ref{fig8} shows an example of four interesting points; if $\mathcal{M}^{\prime}_{i1}=\textbf{true}$, the two sub-problems of $\mathcal{M}^{\prime}_{i3}$ and $(\mathcal{M}^{\prime}_{i2},\mathcal{M}^{\prime}_{i4})$ become independent. Inspired by conditioning techniques for probabilistic databases \cite{KochO08}, and the dynamic generation of optimization units for MapReduce workflows \cite{LimHB12}, we build a reachability graph $RG$ over $\mathcal{M}^{\prime}_i$ to determine a list of \emph{cut sets}. We use single points, composite points of equivalent inputs, and non-overlapping pairs of these as candidates. For each candidate, we determine the points $S_1$ reachable from the roots to the cut set, and the points $S_2$ reachable from the cut set. A cut set $\mat{cs}$ is valid iff $S_1 \cap S_2 = \emptyset$, $S_1 \neq \emptyset$, and $S_2 \neq \emptyset$. Cut sets are then sorted in ascending order of their scores  
\begin{equation2} \label{eq:3}
(2^{\card{\mat{cs}}}-1)/2^{\card{\mat{cs}}} \cdot 2^{\card{\mathcal{M}^{\prime}_{i}}} + 1/2^{\card{\mat{cs}}} \cdot (2^{\card{S_1}} + 2^{\card{S_2}}),
\end{equation2} 
which accounts for the cut set's size and partitioning quality. This order in turn determines the search space layout to which we append all other interesting points. Finally, we materialize the sub-problems for each $\mat{cs}$ in a way that excludes the cut set itself to avoid redundant enumeration. During enumeration, we then probe these cut sets (line~\ref{a3:6}), call \textsc{MPSkipEnum} recursively for their sub-problems (lines~\ref{a3:9}-\ref{a3:10}), and combine the partial plans into the current plan for costing, after which we prune the subspace (line~\ref{a3:20}). 

\textbf{Constraints and Distributed Operations:} We handle the constraints $Z$ via a best-effort prefiltering, where entries that are known to violate constraints---e.g., Row templates with large inputs and violated blocksize constraints---are removed. Remaining violations are assigned infinite costs during enumeration and costing. Similarly, we also use different read bandwidths for inputs of resulting distributed operations to reflect the cost of distributed joins and broadcasts, according to the input sizes of computed cost vectors. 

\section{Experiments}
\label{sec:eval}

Our experiments study the performance characteristics of code generation for linear algebra programs and the optimization of fusion plans. To cover the wide variety of machine learning workloads, we consider (1) several interesting micro-benchmark patterns, (2) dense, sparse, ultra-sparse, and compressed data (of synthetic and real datasets), as well as (3) single-node and large-scale, end-to-end experiments.

\subsection{Experimental Setting}

\textbf{Setup:} We ran all experiments on a 1+6 node cluster consisting of one head node (2x4 Intel E5530\,@\,2.40\,GHz-2.66\,GHz, hyper-threading, $64\gb$ RAM\,@800\,MHz) and six worker nodes (2x6 Intel E5-2440\,@\,2.40\,GHz-2.90\,GHz, hyper-threading, $96\gb$ RAM\,@1.33\,GHz, registered ECC, 12x$2\tb$ disks), 10Gb Ethernet, and CentOS Linux 7.4. The nominal peak memory bandwidth and compute per node are 2x$32\gbs$ from local memory ($47.9\gbs$ measured with the STREAM benchmark \cite{stream}), 2x$12.8\gbs$ over QPI (Quick Path Interconnect), and 2x$115.2\gflops$. We used OpenJDK 1.8.0\_144, Python 2.7.5, Apache Hadoop 2.7.3, and Apache Spark 2.1.0, in yarn-client mode, with 6 executors, 24 cores per executor, $35\gb$ driver memory, $60\gb$ executor memory, and default memory fraction (0.6), which results in an aggregate cluster memory of $6 \cdot 60\gb \cdot 0.6 = 216\gb$. Finally, our framework is integrated in SystemML 0.15\footnote{Our code generation framework is part of Apache SystemML and thus, available open source at \texttt{github.com/apache/systemml}.}.

\textbf{Datasets:} As input data, we use both synthetic and real-world datasets to study different data characteristics, compression, and skew of non-zero values. We created the synthetic data with \texttt{rand} and algorithm-specific data generation scripts. The used real-world datasets are \textbf{Airline78} (years 2007/2008 of the Airline dataset \cite{airline}, $\num{14462943}\times 29$; 0.73; dense), \textbf{Mnist1m}/\textbf{Mnist8m} (scaled versions of the Mnist60k dataset, created with the InfiMNIST data generator \cite{infimnist}, $\num{8100000} \times 784$ and $\num{1012500} \times 784$; 0.25; sparse), \textbf{Netflix} (the Netflix Prize Data \cite{netflix}, $\num{480189}\times\num{17770}$; 0.012; sparse), and \textbf{Amazon} (the books category from the Amazon product review dataset \cite{HeM16,amazon}, $\num{8026324}\times\num{2330066}$; 0.0000012; ultra-sparse). For large-scale experiments, we then scale the data via data generation.

\begin{table}[!t]
\vspace{-0.2cm}
\centering \small \setlength\tabcolsep{3pt}
  \caption{\label{tab:algorithms}ML Algorithms and Configurations (for algorithm details see \texttt{systemml.apache.org/algorithms}).}
  \begin{tabular}{|c|c|c|c|c|c|}
	  \hline
    \textbf{Name} & \textbf{Type} & \textbf{Icpt} & \textbf{$\lambda$} & \textbf{$\epsilon$} & \textbf{MaxIter} \\ 
    \hline
		L2SVM & 2 classes & 0 & $10^{-3}$ & $10^{-12}$ & $20/\infty$\\
		MLogreg & 2/5 classes & 0 & $10^{-3}$ & $10^{-12}$ & $20/10$\\
		GLM & bin.-probit & 0 & $10^{-3}$ & $10^{-12}$ & $20/10$\\
		KMeans & 1 run, $k$=5 & N/A & N/A & $10^{-12}$ & $20$\\
		ALS-CG & rank=20, wL2 & N/A & $10^{-3}$ & $10^{-12}$ & $20/\text{rank}$\\
		AutoEncoder & $\card{\text{batch}}$=512 & N/A & N/A & N/A & $\frac{\text{nrow}(\mat{X})}{\card{\text{batch}}}$ \\
		 & $H_1$=500, $H_2$=2 &&&& \\
    \hline
  \end{tabular}
	\normalsize
	\vspace{0.3cm}
\end{table}

\textbf{ML Algorithms:} To cover the diversity of ML algorithms and their workload characteristics, we conduct end-to-end experiments for six different algorithms from the categories of classification, regression, clustering, matrix factorization, and artificial neural networks. Table~\ref{tab:algorithms} provides an overview of these algorithms and their configurations. The parameters Icpt, $\lambda$, $\epsilon$, and MaxIter refer to the intercept type, the regularization, the convergence tolerance, and the maximum number of outer (and inner) iterations. 

\textbf{Baselines:} As baseline comparisons, we use the following systems with consistent double precision (i.e., \texttt{FP64}) inputs:
\begin{itemize2}
  \item \emph{SystemML 0.15++:} The SystemML (as of Oct'17) baselines are \textbf{Base} (with basic operators), \textbf{Fused} (with hand-coded fused operators, SystemML's default), and \textbf{Gen}. Here, Gen refers to our cost-based optimizer, but we also compare the fuse-all (\textbf{Gen-FA}) and fuse-no-redundancy (\textbf{Gen-FNR}) heuristics.
	\item \emph{Julia 0.6:} As a baseline with LLVM code generation, we use \textbf{Julia} \cite{BezansonEKS17} (without fused operators), and \textbf{JuliaGen} (with fused operators using Julia's dot syntax) \cite{juliagen}. Similar to SystemML, Julia dispatches operations internally to sparse and dense kernels. 
	\item \emph{TensorFlow 1.3:} As a second baseline, we use TensorFlow (\textbf{TF}) \cite{AbadiBCCDDDGIIK16} (without fusion), and \textbf{TFGen}, i.e., TensorFlow XLA \cite{xla}, but only for dense operations due to very limited support for sparse tensors. We built TF from sources with \texttt{-march=native -O3} to enable architecture-specific optimizations and XLA.
\end{itemize2}
Julia and TF call matrix multiplications of native BLAS libraries and Eigen, for both of which we enabled 24 threads. 

\begin{figure*}[!t]
	\centering
	\subfigure[$\text{sum}(\mat{X}\odot\mat{Y}\odot\mat{Z})$, dense]{
	   \label{exp1a}\includegraphics[scale=0.41]{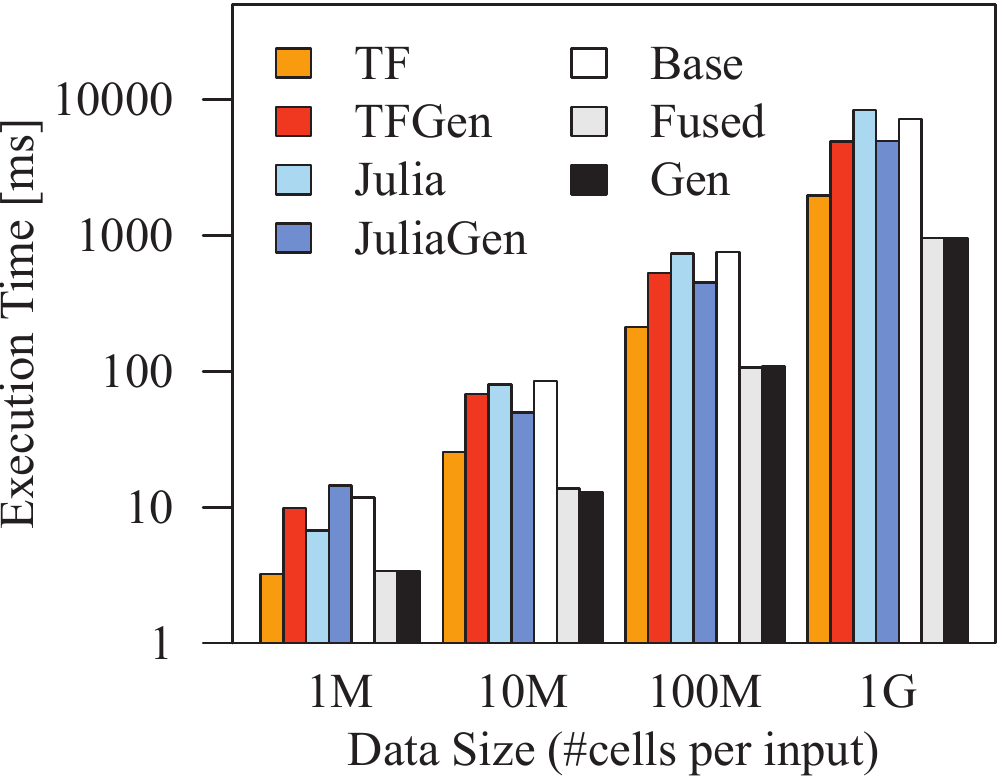}}
	\hfill	
	\subfigure[$\text{sum}(\mat{X}\odot\mat{Y}\odot\mat{Z})$, sparse]{
	   \label{exp1b}\includegraphics[scale=0.41]{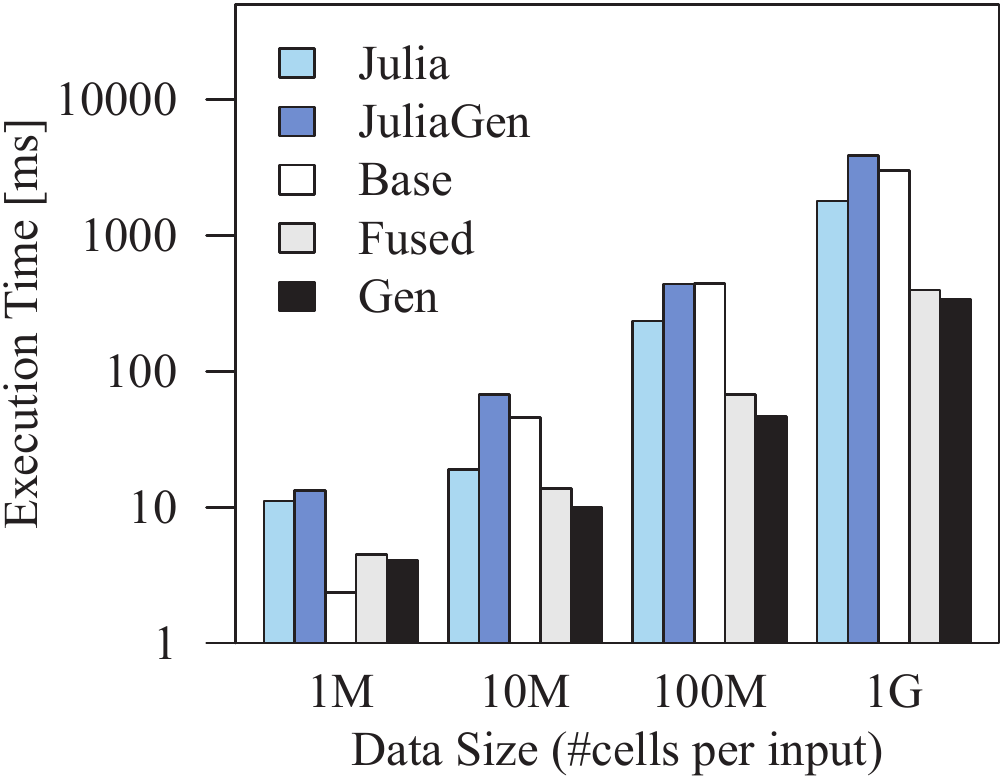}}
	\hfill	
	\subfigure[$\text{sum}(\mat{X}\odot\mat{Y})$, $\text{sum}(\mat{X}\odot\mat{Z})$, d]{
	   \label{exp3a}\includegraphics[scale=0.41]{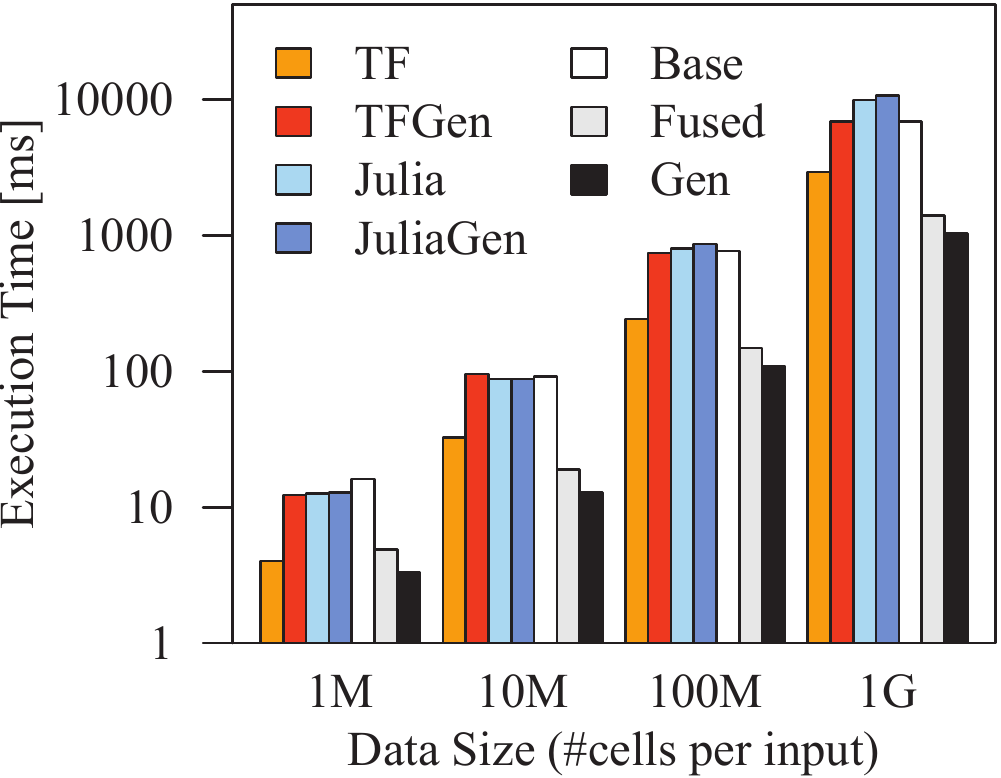}}
	\hfill	
	\subfigure[$\text{sum}(\mat{X}\odot\mat{Y})$, $\text{sum}(\mat{X}\odot\mat{Z})$, s]{
	   \label{exp3b}\includegraphics[scale=0.41]{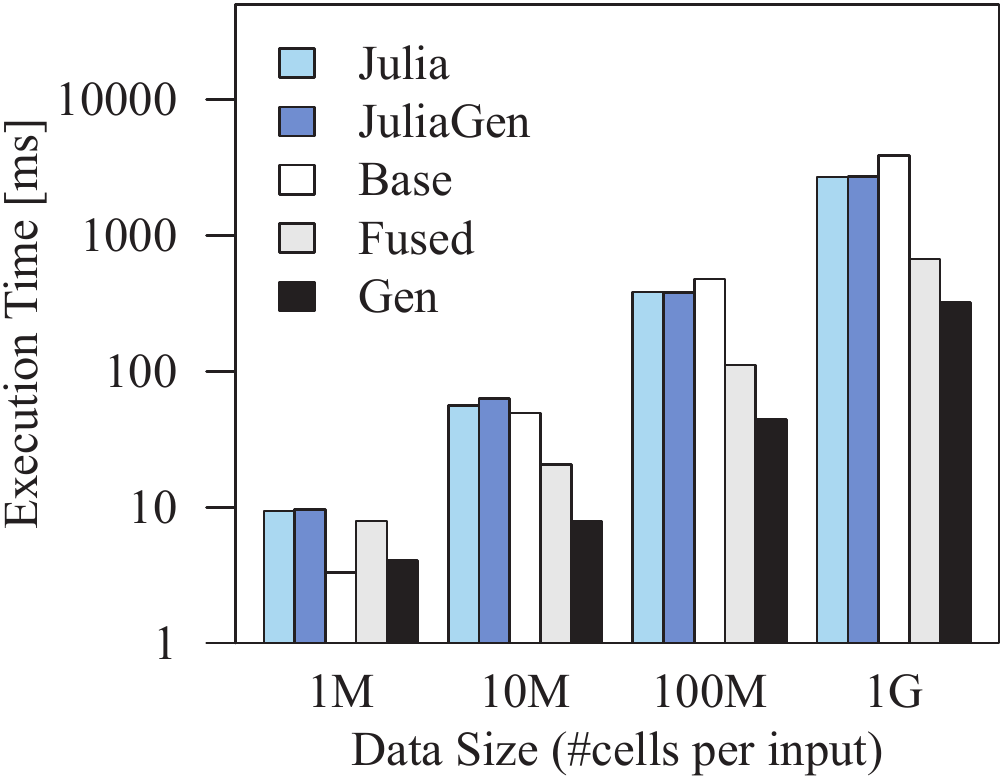}}~\vspace{-0.2cm}~\\
	\subfigure[$\mat{X}^{\top}(\mat{X}\mat{v})$, dense]{
	   \label{exp2a}\includegraphics[scale=0.41]{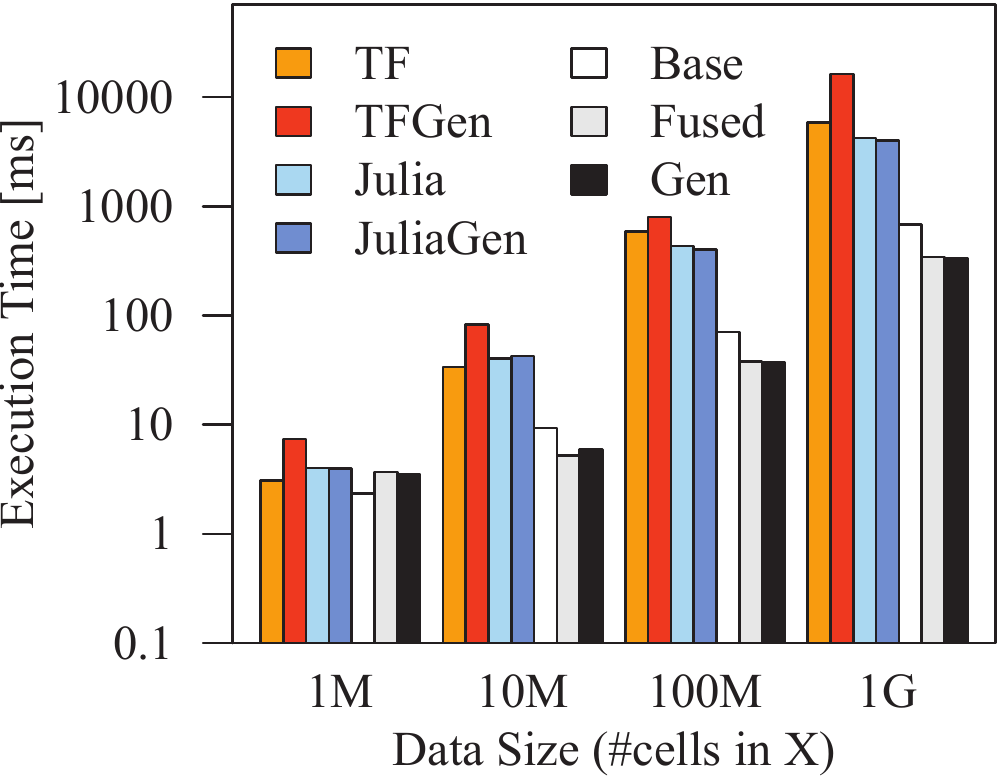}}
	\hfill	
	\subfigure[$\mat{X}^{\top}(\mat{X}\mat{v})$, sparse]{
	   \label{exp2b}\includegraphics[scale=0.41]{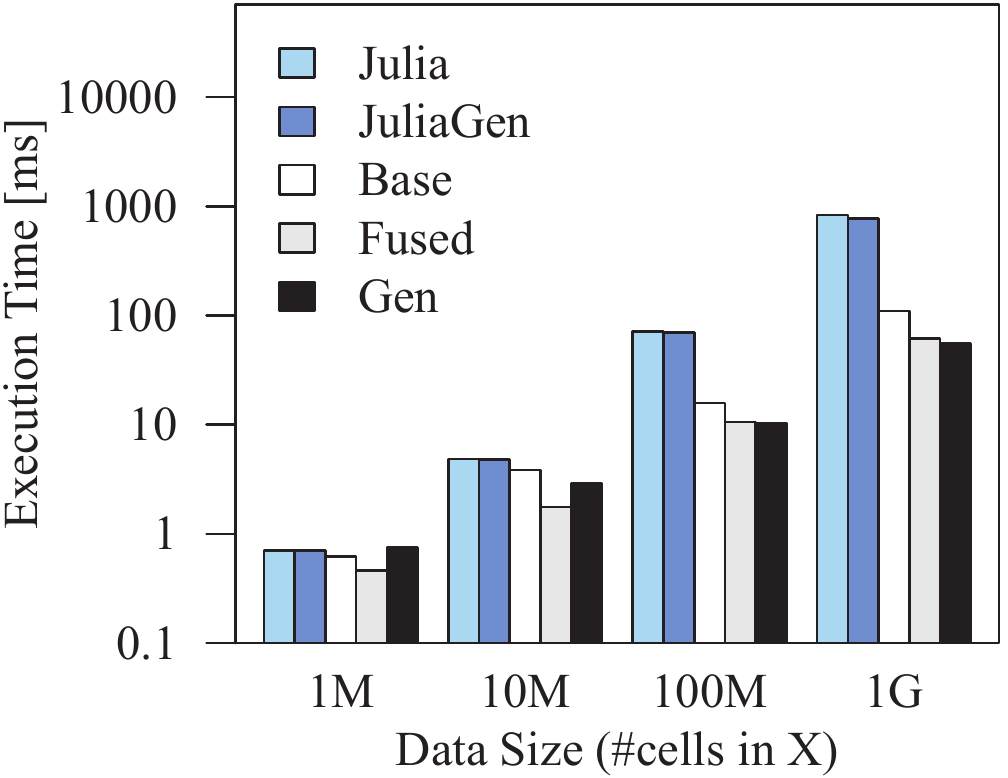}}
	\hfill	
	\subfigure[$\mat{X}^{\top}(\mat{X}\mat{V})$, dense]{
	   \label{exp2c}\includegraphics[scale=0.41]{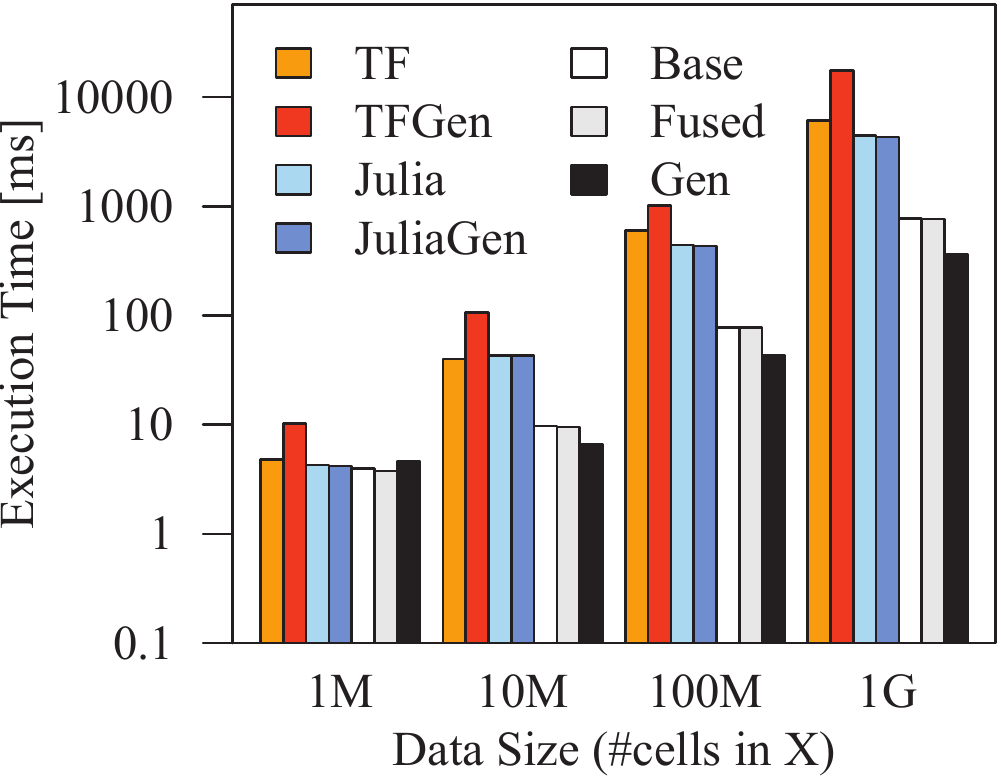}}
	\hfill	
	\subfigure[$\text{sum}(\mat{X} \odot \text{log}(\mat{U} \mat{V}^{\top}+10^{-15}))$]{
	   \label{exp4}\includegraphics[scale=0.41]{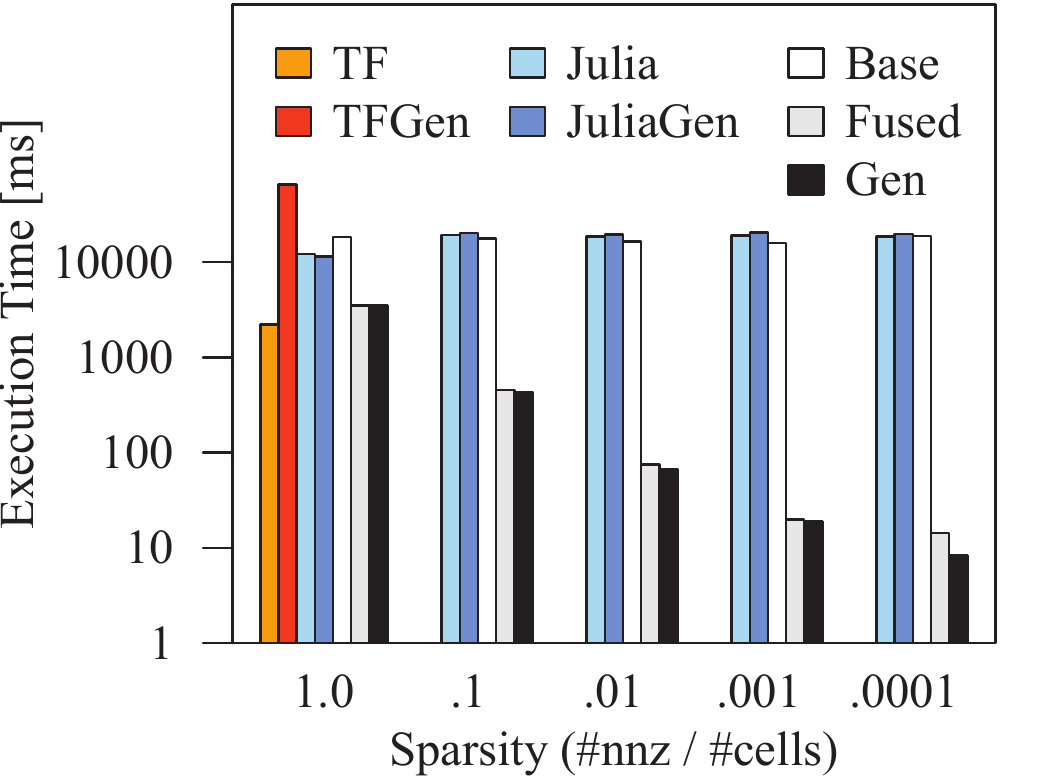}}	
	\vspace{-0.5cm}
  \caption{\label{fig:micro}Operations Performance of Example Patterns (Cell, MAgg, Row, Outer).}
	\vspace{-0.35cm}
\end{figure*}

\subsection{Operations Performance}


In a first set of experiments, we study the multi-threaded performance of our four templates on representative expressions, which have been introduced in Figure~\ref{fig:potential}. These experiments were run on a single worker node, through SystemML's JMLC API (prepared scripts with in-memory inputs), and with the JVM flags \texttt{-Xmx80g -Xms80g -Xmn8g -server}. We used 5 warmup runs for JIT compilation (50 for scenarios with a total input size of $\leq80\mb$) and report the mean runtime of 20 subsequent runs, including recompilation (and thus, cplan construction) overhead.

\textbf{Cell Operations:} Figures~\ref{exp1a} and \ref{exp1b} show the runtimes of Base, Fused, and Gen for $\text{sum}(\mat{X}\odot\mat{Y}\odot\mat{Z})$ over dense and sparse data, compared to Julia and TF. Each input is of size $x\times10^3$ (with sparsity $0.1$ for sparse data), where we vary the number of rows with $x \in \{10^3, 10^4, 10^5, 10^6\}$. For the small $10^3\times10^3$ input (i.e., $8\mb$), Fuse and Gen are only 4x faster because intermediates fit into the L3 cache ($15\mb$). However, as we increase the datasize, Fused and Gen yield performance improvements of an order of magnitude and reach peak single-socket/remote memory bandwidth of $\approx 25\gbs$. In contrast, JuliaGen shows only moderate improvements over Julia because the aggregation is not fused and both, Julia and JuliaGen are single-threaded. TF's multi-threaded operations are competitive for small data because it reuses allocated intermediates. However, Gen is 2.4x faster for larger data because TF still writes intermediates. TFGen shows a consistent slowdown due to single-threaded operations. Furthermore, operator fusion for sparse inputs is much more challenging. In fact, Julia Gen causes---on sparse data---also a slowdown due to sparse lookups. Gen handles such cases more efficiently via stateful iterators under the covers of the stateless \texttt{getValue()} abstraction. 

\textbf{Multi-Aggregate Operations:} Figure~\ref{exp3a} and \ref{exp3b} show the runtimes for the two aggregate operations $\text{sum}(\mat{X}\odot\mat{Y})$ and $\text{sum}(\mat{X}\odot\mat{Z})$ over dense and sparse data as described before. These aggregates qualify as multi-aggregate operation due to their shared input $\mat{X}$. Overall, the performance characteristics are similar to Cell operations with two notable differences. First, the performance of Julia and JuliaGen are identical---except for special cases with different garbage collection behavior---because Julia does neither fuse element-wise operations with aggregations nor consider multi-aggregates. Second, the hand-coded operators of Fused (and similarly the fusion heuristic Gen-FA and Gen-FNR) only apply to $\text{sum}(\mat{X}\odot\mat{Y})$ and $\text{sum}(\mat{X}\odot\mat{Z})$ individually, causing a redundant read of $\mat{X}$. In contrast, Gen compiles a multi-aggregate (with a $2\times1$ output matrix), and in case of sparse data, correctly selects $\mat{X}$ as sparse driver which makes the entire multi-aggregate sparse-safe.

\textbf{Row Operations:} Beside cell-wise operations, row-wise operations are also very common. Figures~\ref{exp2a} and \ref{exp2b} show the runtimes for the matrix-vector multiplication chain $\mat{X}^{\top}(\mat{X}\mat{v})$ over sparse and dense data, where we vary the size of $\mat{X}$ as described before and $\mat{v}$ is a $10^3\times 1$ vector. Julia does not fuse these matrix-vector operations and suffers from single-threaded execution. Despite multi-threaded operations, TF performs even worse because calling Eigen seems to require a copy of $\mat{X}$. TFGen causes again a consistent slowdown. Interestingly, however, with TF's default \texttt{FP32} instead of \texttt{FP64} inputs, TFGen shows a substantial improvement from $5.7\s$ to $0.7\s$ for the 1G scenario. In contrast, Base, Fuse, and Gen achieve peak single-socket/remote memory bandwidth. Fuse and Gen yield a 2x improvement by exploiting temporal row locality, where each row ($8\kb$) fits into L1 cache ($32\kb$). Furthermore, Figure~\ref{exp2c} shows the results of a matrix-matrix multiplication chain $\mat{X}^{\top}(\mat{X}\mat{V})$ over dense data, where $\mat{V}$ is a $10^3\times 2$ matrix. Base and Fused now show equivalent performance because the hand-coded \texttt{mmchain} operator only applies to matrix-vector chains. In contrast, Gen yields a robust 2x performance improvement. 

\textbf{Outer-Product Operations:} Figure~\ref{exp4} reports the runtime of the outer-product operation $\text{sum}(\mat{X} \odot \text{log}(\mat{U} \mat{V}^{\top}+10^{-15}))$, which can exploit sparsity over $\mat{X} \odot$. We fix the size of $\mat{X}$ to $2\cdot 10^4 \times 2\cdot 10^4$ cells, the rank of $\mat{U}$ and $\mat{V}$ to 100, and vary the sparsity of $\mat{X}$ with $\text{sp} \in \{1, 10^{-1}, 10^{-2}, 10^{-3}, 10^{-4}\}$. Base, Julia, and JuliaGen show almost constant runtime, which means Julia does not exploit sparsity. Julia calls a multi-threaded native BLAS matrix multiplication but this expression is largely dominated by the costs for $\text{log}()$, where $\mat{U} \mat{V}^{\top}$ attributes to less than 15\%. For this reason, Base with native BLAS only slightly improved performance by less than a second. For dense data, TF shows very good performance---which is likely due to a different $\text{log}()$ implementation---but does not support sparse operations. In contrast, Fused and Gen achieve, for $\text{sp}=10^{-4}$, a performance improvement of three orders of magnitude and even if $\mat{X}$ is dense, an improvement of $5x$ compared to Base, due to mutli-threaded execution without any intermediates.

\begin{figure}[!b]
	\centering
	\subfigure[Airline78 Dataset (dense)]{
	   \label{exp5a}\includegraphics[scale=0.395]{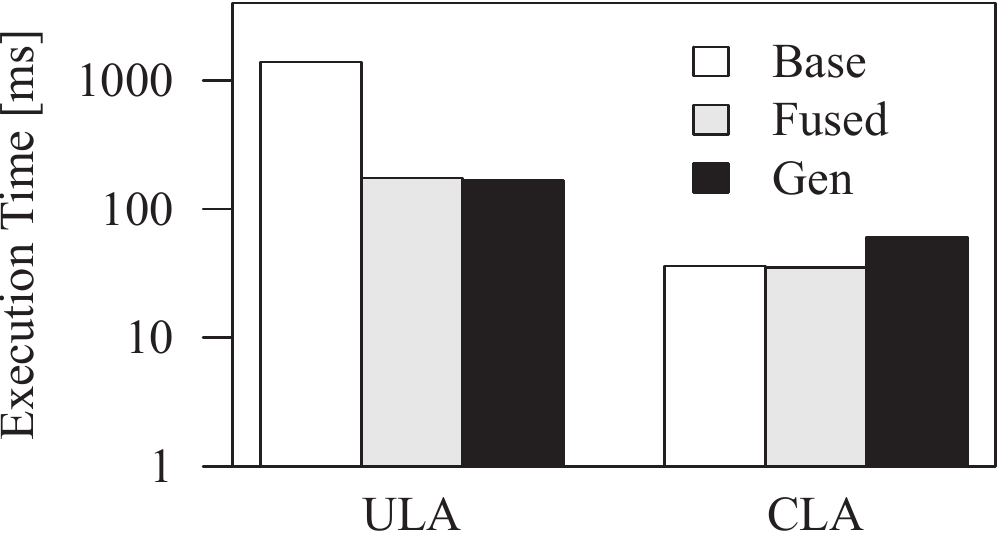}}
	\hfill	
	\subfigure[Mnist8m Dataset (sparse)]{
	   \label{exp5b}\includegraphics[scale=0.395]{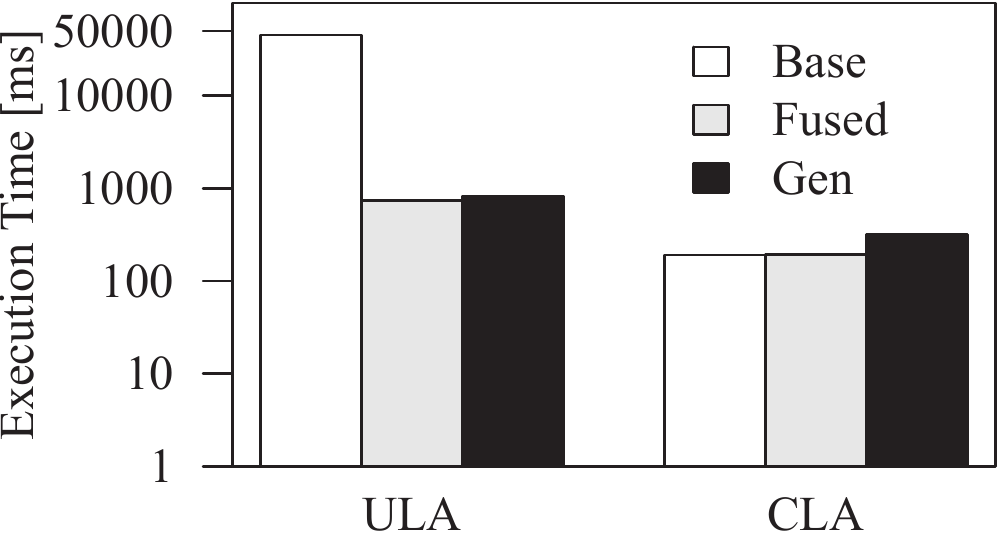}}
	\vspace{-0.5cm}
  \caption{\label{fig:cla}Compressed Operations: $\text{sum}(\mat{X}^2)$.}
	\vspace{-0.1cm}
\end{figure}

\textbf{Compressed Linear Algebra (CLA):} All templates support operations over compressed matrices (column-wise compression, heterogeneous encoding formats, and column co-coding) \cite{ElgoharyBHRR16}. Figure~\ref{fig:cla} shows the runtime of Base, Fused, and Gen for computing the sparse-safe expression $\text{sum}(\mat{X}^2)$ over Airline78 and Mnist8m. For these datasets, CLA achieves compression ratios of $7.44x$ and $7.32x$ over their uncompressed sizes of $3.3\gb$ and $19\gb$. On uncompressed data (ULA), fused operators yield similar speedups as observed for synthetic data because they avoid the expensive materialization of $\mat{X}^2$. On compressed data (CLA), however, Base and Fused show equivalent performance for this special case, because $\mat{X}^2$ is only computed over the dictionary of distinct values with a shallow copy of the compressed data. CLA achieves substantial improvements due to computing the sum via counts per value and reduced memory bandwidth requirements. The template skeletons of Gen exploit similar techniques, by calling---under the conditions of a single input and sparse-safe operations---the generated operator only for distinct values, which achieves performance remarkably close to hand-coded CLA operations.

\begin{figure}[!t]
	\centering
	\subfigure[Default Configuration]{
	   \label{exp6a}\includegraphics[scale=0.395]{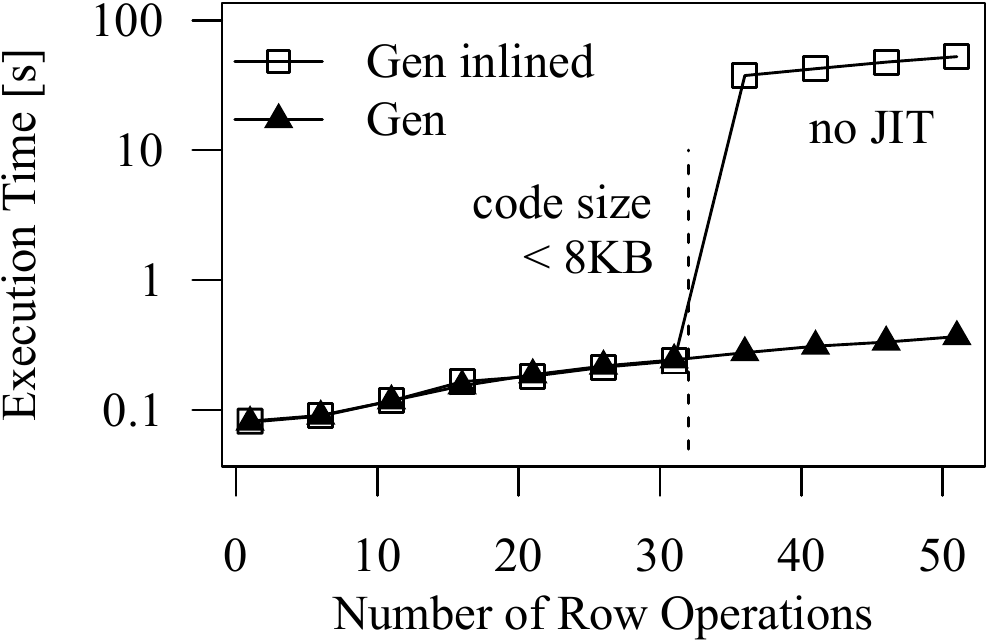}}
	\hfill	
	\subfigure[\scriptsize\texttt{-XX:-DontCompileHugeMethods}]{
	   \label{exp6b}\includegraphics[scale=0.395]{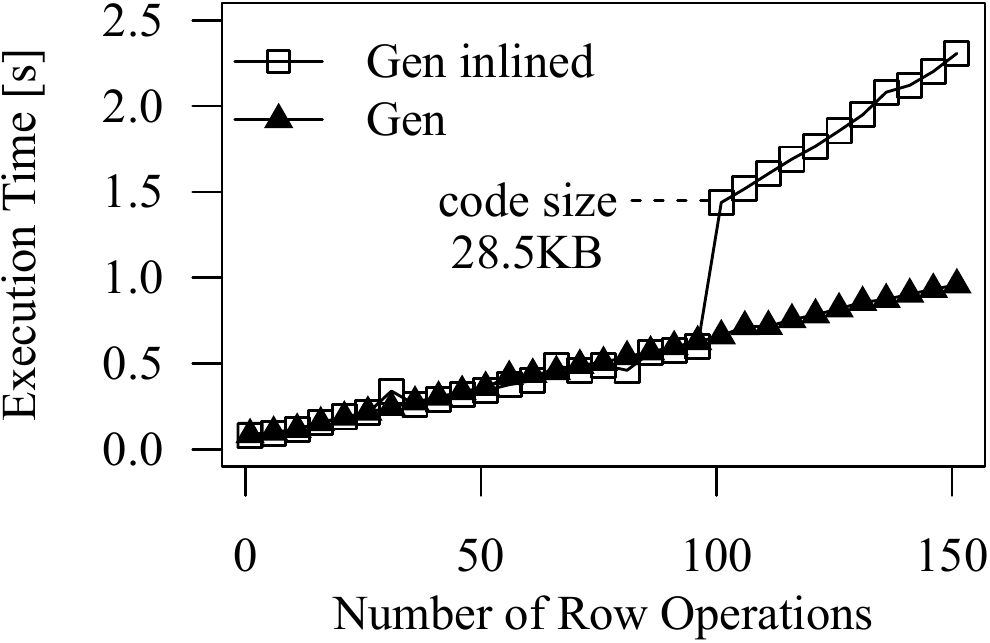}}
	\vspace{-0.35cm}
  \caption{\label{fig:footprint}Impact of Instruction Footprint.}
	\vspace{-0.1cm}
\end{figure}

\textbf{Instruction Footprint:} Separating operator skeletons and vector primitives from the generated operators reduces the instruction footprint. To evaluate its impact, we use $\text{sum}(f(\mat{X}/\text{rowSums}(\mat{X})))$, where we generate $f$ as a sequence of $n$ row operations $\mat{X}\odot i$ and $\mat{X}$ as a dense $10^5\times10^3$ matrix ($800\mb$). Gen uses the vector primitive \texttt{vectMultWrite} (with 8-fold loop unrolling) and---independent of $n$---two vector intermediates per thread. Figure~\ref{exp6a} shows the runtime of Gen and Gen inlined, where the latter inlines the code of \texttt{vectMultWrite}. We set \texttt{-XX:ReservedCodeCache Size=2g} to avoid exceeding the default of $240\mb$. For more than 31 operations, Gen inlined is two orders of magnitude slower because the code size of its \texttt{genexec} method exceeds $8\kb$, which is the JVM threshold for JIT compilation. Figure~\ref{exp6b} reports the runtime with disabled threshold, where both operators show equivalent performance up to 96 operations. However, for 101 and more operations, Gen inlined does no longer fit into the L1 instruction cache ($32\kb$), which leads to a significant performance deterioration. 

\subsection{Compilation Overhead}
\label{sec:overhead}

In a second set of experiments, we investigate the compilation overhead of code generation and optimization. Since the relative overhead decreases with increasing size, we run this over the very small Mnist60k dataset ($60\text{K} \times 784$, sparse). We use again the single worker node setup and report algorithm-level statistics (as the mean of 5 runs, including read times). 

\begin{table}[!b]
\vspace{-0.2cm}
\centering \small \setlength\tabcolsep{4.3pt}
  \caption{\label{tab:overhead}End-to-End Compilation Overhead.}
  \begin{tabular}{|c|c|c|c|c|c|}
	  \hline
    \textbf{Name} & \textbf{Total [s]} & \textbf{\# Compile} & \textbf{Compile [ms]} \\ 
    \hline
		L2SVM & 1.0 & 14/20/12 & \textbf{54} (38)\\
		MLogreg & 3.1 & 429/\num{1580}/26 & \textbf{355} (82)\\
		GLM & 3.1 & \num{3083}/191/57 & \textbf{388} (226)\\
		KMeans & 1.7 & \num{66}/53/18 & \textbf{84} (56)\\
		ALS-CG & 79.7 & \num{1662}/\num{5658}/21 & \textbf{965} (56)\\
		AutoEncoder & 24.2 & 132/\num{2259}/17 & \textbf{452} (54)\\
    \hline
  \end{tabular}
	\normalsize
\end{table}

\textbf{Codegen Statistics:} Table~\ref{tab:overhead} shows the summary code generation statistics---with Gen defaults---for the different algorithms.  These statistics include the execution time, number of compiled plans (optimized HOP DAGs, created CPlans, and compiled Java classes), as well as the compilation overhead (total code generation time, and Java class compilation time). Overall, the overhead is very small---below one second for all algorithms---despite a substantial number of optimized DAGs (up to \num{3083}), constructed CPlans (up to \num{5658}), and compiled operators (up to 57). 

\begin{figure}[!t]
	\centering
	\subfigure[Without Plan Cache]{
	   \label{exp7a}\includegraphics[scale=0.383]{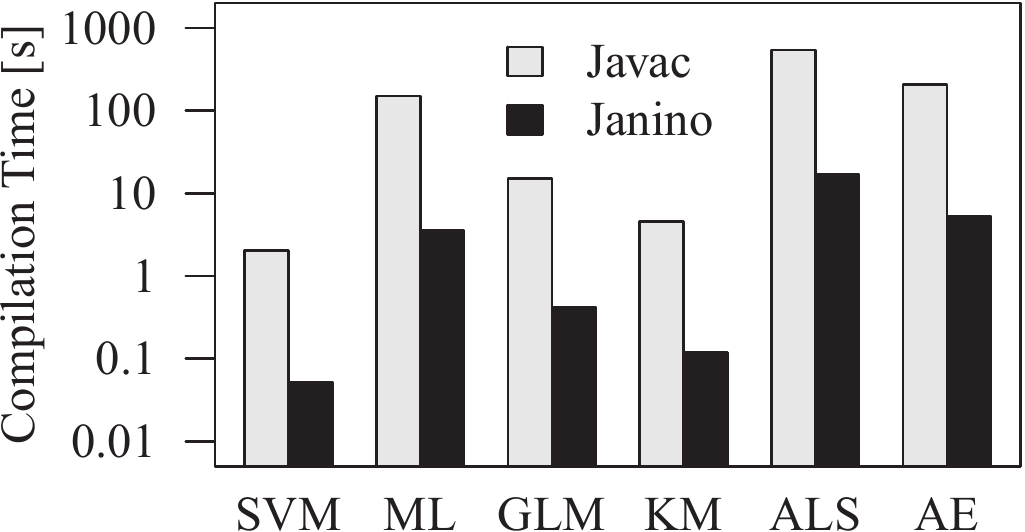}}
	\hfill	
	\subfigure[With Plan Cache]{
	   \label{exp7b}\includegraphics[scale=0.383]{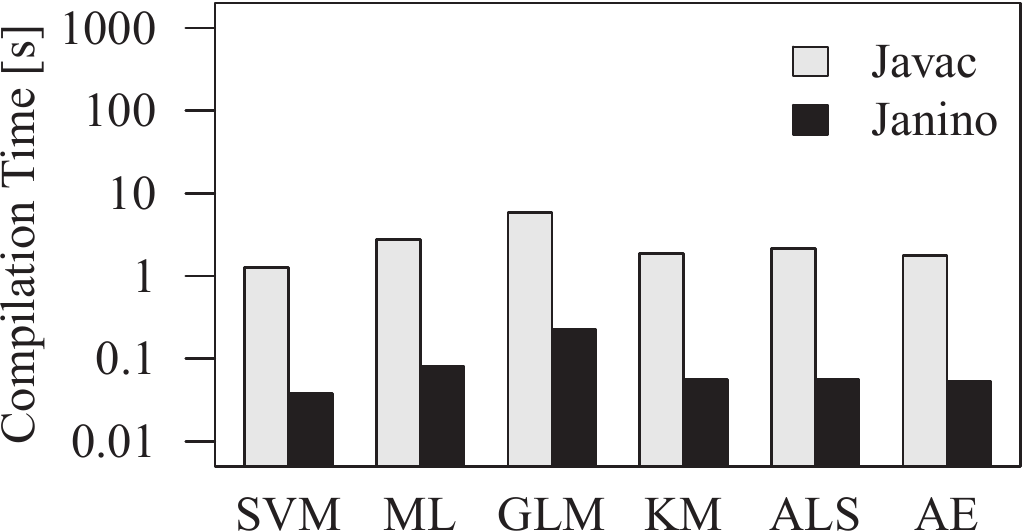}}
	\vspace{-0.5cm}
  \caption{\label{fig:compile}Java Class Compilation and Loading.}
	\vspace{-0.1cm}
\end{figure}

\textbf{Operator Compilation:} By default, Gen uses a plan cache for reusing compiled operators (across DAGs and during dynamic recompilation) as well as the fast \texttt{janino} compiler. Figure~\ref{fig:compile} shows the impact of these components on the compilation overhead. There are two major insights. First, \texttt{janino} consistently improves the performance of class compilation and loading by one and a half orders of magnitude compared to the standard \texttt{javac} compiler. Second, the plan cache also significantly reduces the compilation overhead, especially for algorithms with dynamic recompilation (i.e., MLogreg, GLM, ALS-CG, and AutoEncoder). In detail, the observed plan cache hit rates of the six algorithms are 4/20, \num{1494}/\num{1520}, 88/145, 27/45, \num{5636}/\num{5657}, and \num{2242}/\num{2259}. Besides the reduced compilation overhead, the plan cache also reduces the asynchronous JIT compilation overhead and thus, improves the end-to-end performance. For example, on ALS-CG, the JIT compilation time reduced from $267\s$ to $24\s$, which improved the runtime from $152\s$ to $80\s$.

\begin{figure}[!t]
	\centering
	\includegraphics[scale=0.395]{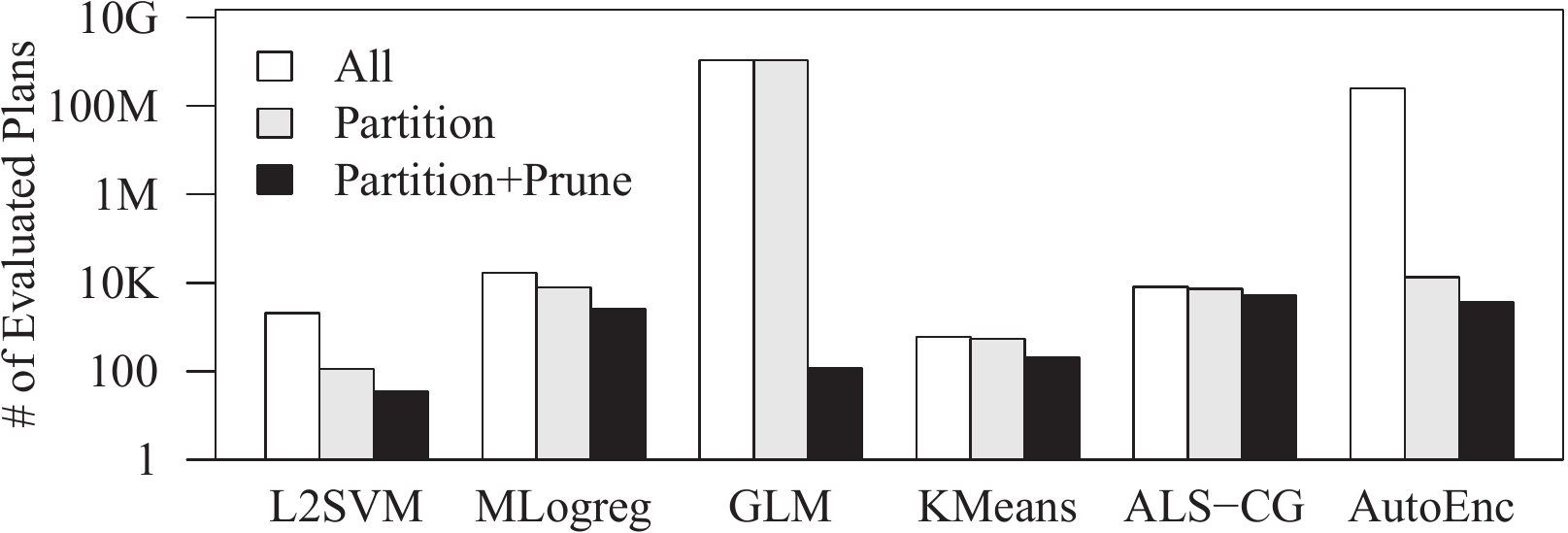}
	\vspace{-0.35cm}
  \caption{\label{fig:enum}Plan Enumeration and Pruning.}
	\vspace{-0.1cm}
\end{figure}

\textbf{Plan Enumeration:} The second major source of overhead is the cost-based plan selection due to its exponential search space. Figure~\ref{fig:enum} shows the total number of evaluated---i.e., costed---plans, for the six algorithms and different configurations without partitioning (all), with partitioning (partition), and with partitioning and both pruning techniques (partition+prune). Overall, none of the algorithms required more than a few thousand plans, for two reasons. First, focusing on interesting points and optimizing partitions---i.e., connected components of fusion plans---independently, is very impactful. For example, the largest DAG of AutoEncoder has 71 operators with partial fusion plans after candidate exploration, which would result in an infeasible number of $2^{71}>10^{21}$ plans. Instead we only consider interesting points. Optimizing partitions independently, further reduces the number of plans by more than four orders of magnitude. Second, the individual pruning techniques, but especially cost-based pruning, are very effective. For example on GLM, pruning reduced the number of evaluated plans by almost seven orders of magnitude.


\subsection{Single-Node End-to-End Experiments}

With an understanding of operation performance and compilation overhead, our third set of experiments studies the end-to-end performance impact of code generation on ML algorithms. Given the results of our micro benchmarks, we restrict this comparison to Base, Fused, and Gen but also include the fusion heuristics Gen-FA, and Gen-FNR in order to evaluate the quality of fusion plans. We report the end-to-end algorithm runtime---invoked through \texttt{spark-submit} with $35\gb$ driver memory---as a mean of 3 runs.

\begin{table}[!t]
\vspace{-0.2cm}
\centering \small \setlength\tabcolsep{5pt}
  \caption{\label{tab:end2end1}Runtime of Data-Int.\,Algorithms [s].}
  \begin{tabular}{|c|c||c|c|c||c|c|}
	  \hline
    \textbf{Name} & \textbf{Data} & \textbf{Base} & \textbf{Fused} & \textbf{Gen} & \textbf{FA} & \textbf{FNR} \\ 
    \hline
	        & $10^6\times10$&6&5&\textbf{3}&3&4\\
		      & $10^7\times10$&36&20&\textbf{6}&7&11\\
		L2SVM & $10^8\times10$&420&228&\textbf{33}&42&93\\
		      \cline{2-7}
		      & Airline78 &123&83&\textbf{20}&24&39\\
					& Mnist8m &180&149&\textbf{103}&123&155\\
		\hline			
			    & $10^6\times10$&8&8&\textbf{5}&5&5\\
		      & $10^7\times10$&59&46&\textbf{13}&14&20\\
		MLogreg & $10^8\times10$&619&481&\textbf{84}&95&209\\
		      \cline{2-7}
				  & Airline78 &184&130&\textbf{46}&49&68\\
					& Mnist8m &494&317&\textbf{243}&288&288\\
		\hline
					& $10^6\times10$&24&24&\textbf{10}&10&14\\
		      & $10^7\times10$&181&176&\textbf{19}&21&52\\
		GLM   & $10^8\times10$&\num{2360}&\num{2266}&\textbf{142}&178&611\\	
		      \cline{2-7}
					& Airline78 &385&338&\textbf{46}&49&105\\
					& Mnist8m &531&392&\textbf{205}&207&291\\
		\hline
					& $10^6\times10$&9&9&\textbf{4}&4&6\\
		      & $10^7\times10$&60&60&\textbf{17}&20&27\\
		KMeans& $10^8\times10$&\num{1543}&\num{1473}&\textbf{123}&153&600\\		
		      \cline{2-7}
					& Airline78 &110&112&\textbf{34}&36&50\\
					& Mnist8m &240&219&\textbf{169}&216&182\\		
    \hline
  \end{tabular}
	\normalsize
\end{table}

\textbf{Data-Intensive Algorithms:} Many traditional ML algorithms are data-intensive, i.e., memory-bandwidth bound. In addition to scans of the feature matrix $\mat{X}$, these algorithms often use many vector and matrix operations, which become a bottleneck for small or very large numbers of features. Accordingly, Table~\ref{tab:end2end1} shows the results for dense input matrices with 10 features, but we also use real datasets. Fused shows only moderate improvements because its patterns are usually limited to two or three operators. Compared to Fused, Gen shows significant end-to-end improvements due to fewer intermediates (which also reduces evictions), fewer scans, and multi-threaded operations with larger scope and thus better utilization. On the $10^8\times10$ ($8\gb$) scenario, we see speedups of 7x, 6x, 16x, and 12x. Regarding heuristics, Gen-FA mostly outperforms Gen-FNR due to fewer intermediates. For this reason, we use Gen-FA as an opening heuristic in Gen. Thanks to cost-based candidate selection, Gen consistently performs about 25\% better than Gen-FA and Gen also automatically adapts plans to the given workload characteristics. For example, Kmeans on Airline78 benefits from full fusion in the inner loop, whereas on Mnist8m a carefully placed intermediate is beneficial.
 
\begin{figure}[!b]
	\centering
	\subfigure[MLogreg \#Classes]{
	   \label{exp9a}\includegraphics[scale=0.395]{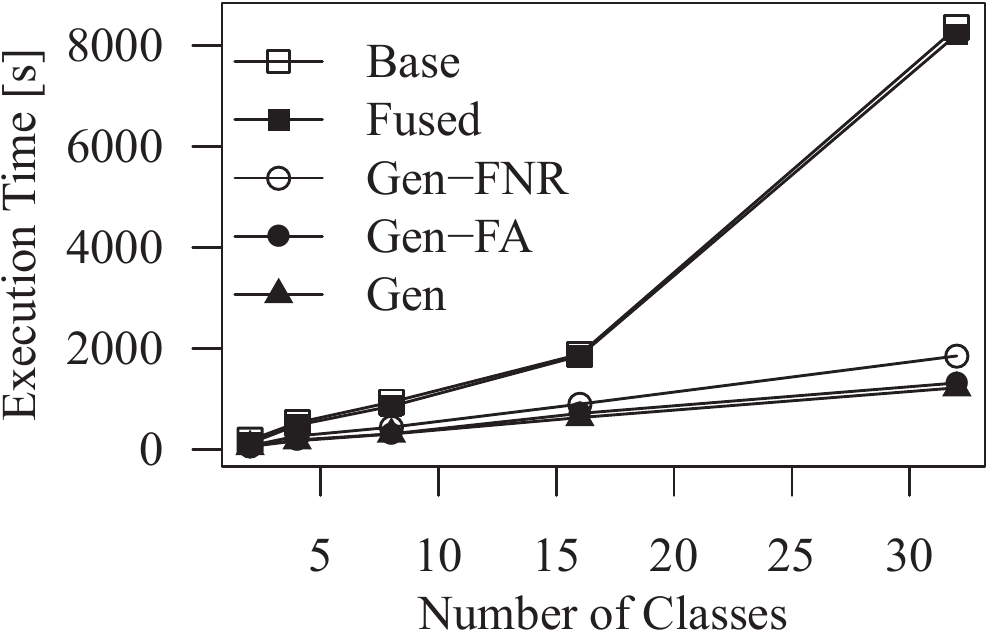}}
	\hfill	
	\subfigure[Kmeans \#Centroids]{
	   \label{exp9b}\includegraphics[scale=0.395]{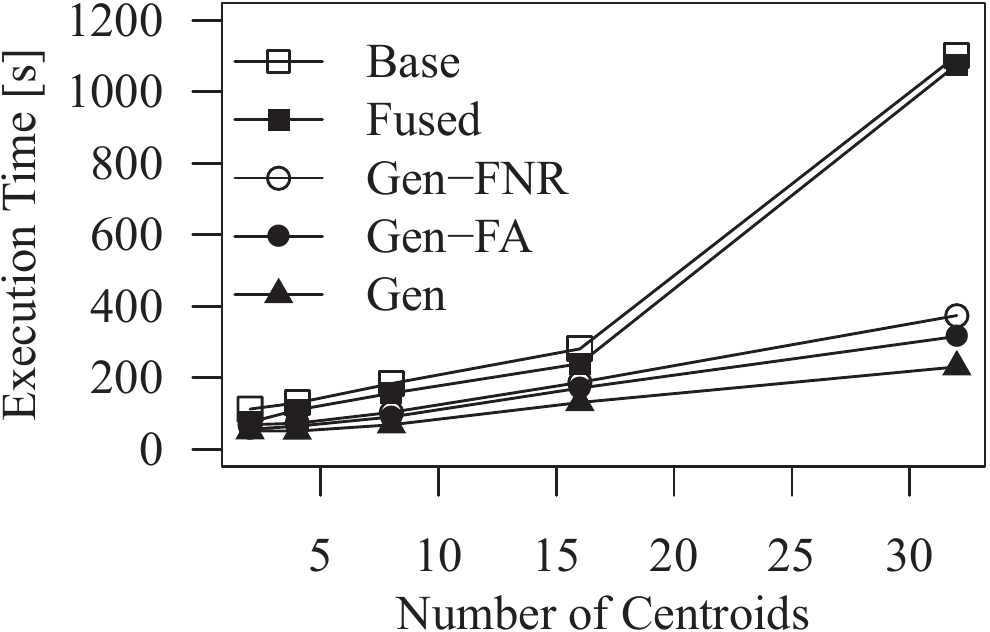}}
	\vspace{-0.35cm}
  \caption{\label{fig:intermediates}Increasing Size of Intermediates.}
	\vspace{-0.2cm}
\end{figure}

\textbf{Hybrid Algorithms:} MLogreg and KMeans are interesting hybrid algorithms, which change from memory-bandwidth- to compute-bound as we increase the number of classes/centroids $k$. Figure~\ref{fig:intermediates} shows the results for an input of size $10^7\times100$ ($8\gb$) and varying $k$. Apart from similar trends as before, there are three insights. First, the runtime of Gen remains almost constant up until $k=8$ because it is still memory-bandwidth-bound, from which onward the runtime increases linearly due to increasing compute workload. Second, $k$ also affects the size of intermediates ($10^7\times k$, i.e., $2.5\gb$ for $k=32$), which causes more evictions for Base and Fused. Third, for the special case of $k=2$, multiple rewrites and fused operators are applied, whereas, Gen shows very robust performance over the entire spectrum.

\begin{table}[!t]
\vspace{-0.2cm}
\centering \small \setlength\tabcolsep{3.4pt}
  \caption{\label{tab:end2end2}Runtime of Compute-Int.\,Algorithms [s].}
  \begin{tabular}{|c|c||c|c|c||c|c|}
	  \hline
    \textbf{Name} & \textbf{Data} & \textbf{Base} & \textbf{Fused} & \textbf{Gen} & \textbf{FA} & \textbf{FNR} \\ 
    \hline
	        & $10^4\times10^4$&528&\textbf{20}&24&233&238\\
		ALS-  & $10^5\times10^4$&\num{24112}&109&\textbf{83}&\num{14180}&\num{13827}\\
		CG& $10^6\times10^4$&N/A&993&\textbf{727}&N/A&N/A\\
		      \cline{2-7}
		      & Netflix &N/A&\num{1063}&\textbf{786}&N/A&N/A\\
					& Amazon  &N/A&\num{18544}&\textbf{\num{8552}}&N/A&N/A\\
		\hline			
			      & $10^3\times10^4$ &8&9&\textbf{7}&7&7\\
		Auto- & $10^4\times10^4$ &51&48&\textbf{32}&33&35\\
		Encoder & $10^5\times10^4$ &615&560&291&\textbf{281}&287\\
		      \cline{2-7}
				  & Mnist1m &597&562&381&381&\textbf{367}\\	
    \hline
  \end{tabular}
	\normalsize
\end{table}

\textbf{Compute-Intensive Algorithms:} We also study the compute-intensive algorithms ALS-CG for matrix factorization and AutoEncoder for dimensionality reduction. Table~\ref{tab:end2end2} shows the results of ALS-CG on sparse data ($0.01$), AutoEncoder on dense data, as well as different real sparse and ultra-sparse datasets. For ALS-CG, Fused and Gen show improvements of multiple orders of magnitude due to sparsity exploitation in the update rules and loss computations. Gen outperforms Fused due to less evictions as the dimensionality increases. The fusion heuristics fail to find good plans for the update rules due to an overlapping Row template that destroys the sparse-safe Outer template. Hence, Base, Gen-FA, and Gen-FNR are not applicable for larger datasets. Even for AutoEncoder, Gen and the fusion heuristics show a solid 2x runtime improvement, despite the used mini-batch algorithm and thus, small intermediates, as well as many compute-intensive matrix-matrix multiplications. Here, the fusion heuristics perform slightly better than Gen though, due to less optimization and JIT compilation overhead.

\begin{table}[!b]
\vspace{-0.3cm}
\centering \small \setlength\tabcolsep{3.4pt}
  \caption{\label{tab:end2end3}Runtime of Distributed Algorithms [s].}
  \begin{tabular}{|c|c||c|c|c||c|c|}
	  \hline
    \textbf{Name} & \textbf{Data} & \textbf{Base} & \textbf{Fused} & \textbf{Gen} & \textbf{FA} & \textbf{FNR} \\ 
    \hline
	        & D200m &\num{1086}&\num{712}&\textbf{335}&\num{1642}&862\\
		L2SVM & S200m & \num{1381}&\num{837}&\textbf{356}&\num{2361}&533\\
		      \cline{2-7}
					& Mnist80m &\num{1567}&\num{1014}&\textbf{577}&\num{1154}&933\\
		\hline			
			    & D200m &\num{4878}&\num{4504}&\textbf{\num{2817}}&\num{7965}&\num{7805}\\
		MLogreg & S200m &\num{4351}&\num{3746}&\textbf{\num{3171}}&\num{4043}&\num{4302}\\
		      \cline{2-7}
					& Mnist80m &\num{5343}&\num{4373}&\textbf{\num{3412}}&\num{8852}&\num{10578}\\
		\hline
					& D200m &\num{13925}&\num{13435}&\textbf{\num{2199}}&\num{2323}&\num{4074}\\
		GLM   & S200m &\num{13598}&\num{12665}&\textbf{\num{2707}}&\num{3646}&\num{4878}\\	
		      \cline{2-7}
					& Mnist80m &\num{6202}&\num{3602}&\textbf{\num{1308}}&\num{1443}&\num{1772}\\
		\hline
					& D200m&\num{5615}&\num{5605}&\textbf{308}&\num{325}&\num{7090}\\
		KMeans& S200m&\num{5421}&\num{5334}&\textbf{248}&\num{259}&\num{6573}\\
		      \cline{2-7}
					& Mnist80m &\num{2174}&\num{2147}&\textbf{395}&\num{412}&\num{4181}\\		
    \hline
  \end{tabular}
	\normalsize
	\vspace{-0.1cm}
\end{table}

\subsection{Large-Scale End-to-End Experiments}

Finally, we also study large-scale (i.e., distributed) algorithms. We use three datasets: D200m ($200\text{M}\times100$, dense, $160\gb$), S200m ($200\text{M}\times10^3$, 0.05, sparse, $121\gb$), Mnist80m ($81\text{M} \times 784$, 0.25, sparse, $204\gb$), which all fit in aggregate memory ($216\gb$), and we report the end-to-end runtime, with $35\gb$ driver, as a mean of 3 runs in Table~\ref{tab:end2end3}.

\textbf{Distributed Algorithms:} Overall, Gen shows again substantial improvements compared to Fused (by up to 21x for KMeans). However, unlike in the single-node experiments, the fusion heuristics show brittle performance characteristics. For example, Gen-FA even leads to slowdowns on L2SVM and MLogreg. This effect is due to eager fusion of vector operations---that could be executed at the driver---into distributed operations over large inputs. In a distributed environment, these additional vector inputs cause unnecessary broadcast overhead and partial evictions of broadcasts from aggregate memory. In contrast, Gen creates good plans by reasoning about template switches and broadcast costs.

\section{Related Work}
\label{sec:rwork}

We review work from query compilation, loop and operator fusion, and the optimization of DAGs and fusion plans.

\textbf{Query Compilation:} Already System~R compiled SQL statements---for repetitive transactions---into machine code \cite{ChamberlinABGKLLMPPSSSTWY81,ChamberlinAKLMPSSSWY81}, but compilation was later abandoned due to maintenance and debugging issues \cite{RaoPML06}. Motivated by the trend toward in-memory databases, query compilation was then reconsidered by JAMDB \cite{RaoPML06}, HIQUE \cite{KrikellasVC10}, DBToaster \cite{KennedyAK11}, and HyPer \cite{Neumann11}. Kennedy et al. introduced the compilation of incremental view maintenance programs in DBToaster \cite{KennedyAK11}, whereas Neumann made a case for LLVM-based query compilation in HyPer to support ad-hoc queries with low compilation overhead \cite{Neumann11}. LegoBase \cite{KlonatosKRC14} and DBLAB/L \cite{ShaikhhaKPBD016} focused on a modular compilation chain to exploit both relational and compiler optimizations. Several systems also include restricted ML workloads into query compilation. Examples are the compilation of UDF-centric workflows in Tupleware \cite{CrottyGDKBCZ15}, Lambda expressions in Hyper \cite{PassingTHLSGK017}, the LLVM-based compilation of Java UDFs \cite{Rosenfeld17}, and query compilation with UDFs in Flare \cite{EssertelTDBOR17}. Lang et al. explored the integration with scans over compressed blocks in Hyper \cite{LangMFB0K16}, which---similar to our Row template over compressed matrices---extracts tuples to temporary storage. Menon et al. further introduced the notion of relaxed operator fusion to reason about temporary materialization in Peloton \cite{Menon17}. Additional directions are compiler abstractions for different hardware backends in Voodoo \cite{PirkMZM16}, and compiled data access over heterogeneous formats in Proteus \cite{KarpathiotakisA16}. Meanwhile, query compilation is heavily used in many modern database systems such as Hyper \cite{0001L14}, Impala \cite{Wanderman-MilneL14}, Hekaton \cite{FreedmanIL14}, MemSQL \cite{memsql}, Tupleware \cite{CrottyGDKBCZ15}, Peloton \cite{PavloAALLMMMPQS17}, and SparkSQL \cite{ArmbrustXLHLBMK15}. However, most of these systems do not handle DAGs, linear algebra operations, or the challenges of sparsity exploitation.

\textbf{Loop and Operator Fusion:} 
Loop fusion, tiling and distribution \cite{AllenK2001,KennedyM93} aim at merging multiple loops into combined loops and vice versa---without introducing redundancy or loop-carried dependencies---to improve locality, parallelism, or memory requirements. Existing work typically relies on the affine \cite{LimL97} or polyhedral \cite{PouchetBBCRSV11} models to build an inter-loop dependency graph \cite{AllenK2001}. Since loop fusion is known to be NP-complete \cite{Darte00,KennedyM93}, typically greedy \cite{Kennedy01} or heuristic \cite{MehtaLY14} methods are used. Also, loop fusion usually only considers dense data access. 
Recent research aims at specialized IRs for staged transformations---which does allow sparsity exploitation for restricted cases of unary operations---\cite{RompfSABJLJOO13}, normalization of comprehensions in Emma \cite{AlexandrovKKSTK15}, distributed applications on heterogeneous hardware \cite{BrownLRSSAO16}, and cross-library optimization in Weld \cite{abs-1709-06416,PalkarTSSAZ17}.
In ML systems, operator fusion aims at merging multiple operations over matrices or tensors into fused operations. In contrast to loop fusion, the dependencies are implicitly given by the data flow graph and operation semantics \cite{BelterJKS09}. SystemML uses rewrites to identify special operator patterns, and replaces them with hand-coded local or distributed fused operators \cite{AshariTBRCKS15,BoehmDEEMPRRSST16,HuangBTRTR15}. Other systems like Cumulon \cite{HuangB013} and MatFast \cite{YuTAMAO17} use more generic masked and folded binary operators to exploit sparsity across matrix multiplications and element-wise operations. Automatic operator fusion addresses the limitations of these approaches. BTO \cite{BelterJKS09} introduced a refine-and-optimize approach for fusing BLAS Level 1/2 operations in local linear algebra kernels, whereas OptiML \cite{SujeethLBRCWAOO11} provided operator fusion for both CPU and GPUs. Tupleware \cite{CrottyGDKBCZ15,CrottyGDKCZ15} and Kasen \cite{ZhangWCMZ16} later introduced operator fusion for distributed programs. In contrast, SystemML-SPOOF \cite{ElgamalLBETRS17} provides operator fusion for local and distributed operations, as well as sparsity-exploitation across operations. Additionally, Sparso \cite{RongPXAS16} introduced a framework to discover, propagate and use context in sparse linear algebra programs. Meanwhile, operator fusion and code generation are being integrated into many systems in practice. Examples are SystemML, TensorFlow XLA \cite{AbadiBCCDDDGIIK16,xla}, Intel Nervana Graph \cite{nervana}, NVIDIA TensorRT \cite{tensorrt}, and Julia \cite{BezansonEKS17,juliagen}. However, most of these systems rely on fusion heuristics or manual declaration.

\textbf{Optimizing DAGs and Fusion Plans:} Operator DAGs are ubiquitous in ML workloads, which is challenging because the optimal solution might not consist of optimal solutions to its subproblems.
Neumann pioneered the work on generating optimal DAG-structured query plans \cite{Neumann2005,NeumannM09}, while others heuristically share CSEs via materialized views \cite{Neumann2005,SilvaLZ12,ZhouLFL07} or common operators \cite{ArumugamDJPP10,CandeaPV09,GiannikisMAK14}. Recent work further introduced a greedy algorithm with guaranteed approximation factor \cite{Kathuria017}. Sideways information passing such as semi-join reductions \cite{BernsteinC81}, magic sets \cite{BancilhonMSU86}, bypass plans for disjunctive queries \cite{SteinbrunnPMK95}, or adaptive information passing \cite{IvesT08,NeumannW09} also deal with DAGs, but none of these techniques are integrated with query compilation.
Although most ML systems have compiler and runtime support for DAGs, their rewrite systems---such as SystemML's static and dynamic rewrites \cite{BoehmBERRSTT14}---also deal with CSEs in a heuristic manner.
Similarly, the literature on optimizing fusion plans is very sparse. Frameworks such as OptiML \cite{SujeethLBRCWAOO11}, Emma \cite{AlexandrovKKSTK15}, Kasen \cite{ZhangWCMZ16}, Voodoo \cite{PirkMZM16}, SystemML-SPOOF \cite{ElgamalLBETRS17}, Weld \cite{PalkarTSSAZ17}, and TensorFlow XLA \cite{AbadiBCCDDDGIIK16,xla} all use fusion heuristics, which misses significant opportunities. Tupleware \cite{CrottyGDKBCZ15} combines heuristics and cost-based decisions for micro-optimizations such as predication and loop tiling. In contrast, BTO \cite{BelterJKS09} uses a greedy algorithm with $k$ current plans that repeatedly applies transformation rules for refinement and optimization, using an analytical cost model. In comparison to our optimization framework, these systems do not ensure optimality of fusion plans and do not exploit sparsity across operations.

\section{Conclusions}

To summarize, we introduced a practical, cost-based optimization framework for operator fusion plans over DAGs of linear algebra operations. This framework includes, (1) an effective compiler and runtime integration for large-scale ML systems, as well as (2) novel algorithms for the efficient exploration of valid fusion candidates, and the cost-based enumeration of fusion plans. Our experiments show that optimized fusion plans match the performance of hand-coded fused operators, and lead---due to their generality and cost-awareness---to significant end-to-end improvements compared to hand-coded operators and fusion heuristics.
In conclusion, we believe that automatic operator fusion and the optimization of fusion plans is a corner-stone of future declarative, large-scale ML systems. The major benefits are its high performance impact, the reduced development effort, and its broad applicability regarding a wide variety of ML algorithms, dense, sparse, and compressed data, as well as local and distributed operations.
Interesting future work includes---as outlined in the SPOOF vision \cite{ElgamalLBETRS17}---the holistic optimization of fusion plans and simplification rewrites, the inclusion of additional classes of operations, as well as code generation for heterogeneous hardware including GPUs.

\balance
\small

\normalsize

\end{document}